\documentclass[aps,
               pra,
               twocolumn,
               amsmath,
               amssymb,
               superscriptaddress,
               reprint,
               10pt
              ]{revtex4-2}

\usepackage{url}
\usepackage{txfonts}
\usepackage{amsmath, amssymb}

\usepackage{pstricks}
\usepackage{graphicx}
\usepackage{tikz}
\usepackage[T1]{fontenc}
\usepackage{color}
\usepackage[colorlinks=true,breaklinks=true,allcolors=blue]{hyperref}

\usepackage{tabularx}
\usepackage{multirow}
\usepackage{array}
\newcolumntype{L}[1]{>{\raggedright\let\newline\\\arraybackslash\hspace{0pt}}m{#1}}
\newcolumntype{C}[1]{>{\centering\let\newline\\\arraybackslash\hspace{0pt}}m{#1}}
\newcolumntype{R}[1]{>{\raggedleft\let\newline\\\arraybackslash\hspace{0pt}}m{#1}}

\usepackage{physics}
\usepackage{xparse}
\usepackage[version=4]{mhchem}
\usepackage{lipsum}
\usepackage[normalem]{ulem}

\renewcommand{\i}{\mathrm{i}}
\newcommand{\e}{\mathrm{e}}

\newcommand{\eq}[1]{(\ref{eq:#1})}
\newcommand{\Eq}[1]{Eq.\,\eqref{eq:#1}}

\newcommand{\Fig}[1]{Fig.~\ref{fig:#1}}
\newcommand{\fig}[1]{\ref{fig:#1}}

\newcommand{\App}[1]{App.~\ref{app:#1}}

\renewcommand*\exp[1]{\mathrm{exp} \qty( #1 )}

\newcommand*\aho{a_\mathrm{ho}}

\newcommand*\xih{\xi}

\renewcommand*\i{\mathrm{i}}

\makeatletter
\let\cat@comma@active\@empty
\makeatother

\begin{document}

%==============================================================================
%==============================================================================
\title{Decaying superfluid turbulence near an anomalous non-thermal fixed point}

\author{Niklas Rasch}
\email{niklas.rasch@kip.uni-heidelberg.de}
\affiliation{Kirchhoff-Institut f\"ur Physik, 
	Ruprecht-Karls-Universit\"at Heidelberg,
	Im Neuenheimer Feld 227, 
	69120 Heidelberg, Germany}

\author{Thomas Gasenzer}
\email{t.gasenzer@uni-heidelberg.de}
\affiliation{Kirchhoff-Institut f\"ur Physik, 
             Ruprecht-Karls-Universit\"at Heidelberg,
             Im Neuenheimer Feld 227, 
             69120 Heidelberg, Germany}
\affiliation{Institut f\"ur Theoretische Physik, 
			Ruprecht-Karls-Universit\"at Heidelberg,
			Philosophenweg 16, 69120 Heidelberg, Germany}

\date{\today}

%==============================================================================
%==============================================================================
\begin{abstract}
	We investigate anomalously slow coarsening in a dilute two-dimensional (2d) superfluid closed with respect to particle and energy exchange with the environment. 
	The dynamics is demonstrated to be closely connected to both, a non-thermal fixed point (NTFP) in a far-from-equilibrium quantum system, and to Kraichnan-Kolmogorov turbulence.
	During a universal dynamical regime associated with an anomalous NTFP, vortex dynamics are understood to be governed by three-vortex collisions that trigger vortex-antivortex annihilation events, leading to a subdiffusive decay of the vortex density and thus  growth of the characteristic inter-defect length scale, \(\ell_\text{v}\sim t^{\,\beta}\) with \(\beta\approx1/5\).
	It is found that, during the same time when this power law in time is seen, the moments of the superfluid velocity circulation $\Gamma$ around an area of spatial extent $r$ exhibit power-law scaling $\Gamma^{2}(r)\sim r^{8/3}$, in agreement with Kraichnan-Kolmogorov predictions for an inverse energy cascade in the inertial range, in a driven-open setting.
	Moreover, in high-order moments, intermittent deviations from linear scaling $\Gamma^{2p}(r)\sim [\Gamma^{2}(r)]^{p}$ are observed that are consistent with bifractal intermittency corrections previously measured in fully developed classical turbulence.
	These results establish a quantitative link between decaying quantum turbulence in a closed superfluid and universal dynamics near a non-thermal fixed point.
	Notably, the subdiffusive decay exponent $\beta\approx1/5$ deviates significantly from values reported for classical systems.
\end{abstract}

\maketitle

%==============================================================================
%==============================================================================
\emph{Introduction}.---%
Fluid turbulence has been among the first systems used to study spatial and temporal scaling phenomena in nature \cite{Frisch1995a,Kolmogorov1941a.RSPSA.434.9,Kolmogorov1941b.RussMath20thC.323,Kolmogorov1941c.RSPSA.434.15,Obukhov1941a,Obukhov1941b,Bodenschatz2011a}.
In three spatial dimensions, driven, fully developed turbulence has been characterized as a direct cascade, in which kinetic energy is transferred from large to small spatial scales through an inertial range, establishing the stationary spectral distribution $E(k)\sim k^{-5/3}$.
In contrast, restricting the dynamics to a (quasi-)2d geometry, the energy is transported in the opposite direction as large-scale eddies are formed from smaller ones, which forms an inverse cascade, giving rise to the same stationary algebraic momentum distribution, $E(k)\sim k^{-5/3}$ \cite{Kraichnan1967a,Leith1968a.PhysFl11.671,Batchelor1969a.PhysFl12.II-233}.

On scales where quantum statistics and coherence prevail, superfluid or quantum turbulence emerges in inviscid superfluids, including ${}^4 \mathrm{He}$ and dilute Bose-Einstein condensates \cite{Volovik2004a,Tsubota2008a,Barenghi2014a.PNAS111.4647,Tsatsos2016a}.
The elementary constituents of the flow are topologically protected quantum vortices, carrying point-like vorticity with discrete values $qh/m$, $q\in\mathbb{Z}$.
They thus differ from classical eddies in their microscopic properties, while on large scales certain characteristics of classical fluid turbulence, like the inverse energy cascade (IEC) in two dimensions, have been recovered in superfluids \cite{Johnstone2019a,Gauthier2019a,Sachkou2019a.Science366.1480}.

While turbulence is commonly studied in fully developed turbulence -- where external driving injects energy at the same rate at which it is dissipated at small scales, e.g., into thermal excitations -- other scenarios of interest include the buildup and decay of a turbulent cascade.
Of particular relevance in this regard is the evolution within a closed system, i.e., without energy and particle gain and loss.
The decay of turbulence has been discussed for classical 3d \cite{George1992a.PhysFl4.1492,Eyink2000a.PhysFl12.477} and 2d turbulence
\cite{Carnevale1991a.PhysRevLett.66.2735,
Matthaeus1991a.PhysicaD51.531,
Tabeling1991a.PhysRevLett.67.3772,
Carnevale1992a.PhysFlA.6.1314.McWilliams:endstate,
Cardoso1994a.PhysRevE.49.454,
Bracco2000a.PhysFl12.2931,
Iwayama2002a,
Danilov2002a.PhysRevE.65.036316,
Bokhoven2007a.PhysFl19.046601,
Dritschel2008a.PhysRevLett.101.094501,
Sire2010a.PhysRevE.84.056317,
Mininni2013a.PhysRevE.87.033002,
Fang2017a.PhysFl29.111105}
as well as for quantum systems
\cite{Chu2001a,
Walmsley2008aNoArxiv,
Skrbek2012a.PhysFl24.011301,
Baggaley2012b.PhysRevB.85.060501,
Forrester2013a.PhysRevLett.110.165303,
Zmeev2015a}.
Both buildup and decaying (wave) turbulent flows can govern condensate formation 
\cite{Svistunov1991a,
Kagan1992a,
Kagan1994a,
Kagan1995a,
Semikoz1995a.PhysRevLett.74.3093,
Semikoz1997a}.
In 2d, studies of decaying quantum turbulence have focused on Onsager-Kraichnan condensation of vortex clusters and the loss of vortex-antivortex pairs 
\cite{Neely2013a.PhysRevLett.111.235301,
Billam2014a.PhysRevLett.112.145301,
Kwon2014a.PhysRevA.90.063627,
Stagg2015a.PhysRevA.91.013612,
Cidrim2016a.PhysRevA.93.033651,
Groszek2016a.PhysRevA.93.043614,
Seo2017aNatSciRept7.4587S,
Baggaley2018a.PhysRevA.97.033601,
Mueller2020a,
Kanai2025a,
Novotny2025a}, 
the direct enstrophy cascade 
\cite{Forrester2014a.JPhysConfSer568.012031,
Reeves2017a.PhysRevLett.119.184502,
Forrester2020a,Williams2022a}, 
and finite temperatures 
\cite{Groszek2020a.SciPostPhys.8.3.039,
Groszek2021a.PhysRevResearch.3.013212}.
The latter work examined the scaling evolution of single-particle spectra, vortex number, and inter-defect distance, characterizing the vortex decay and scaling as universal dynamics near a non-thermal fixed point (NTFP)
\cite{Berges2008a,
Berges2008b,
Scheppach2009a,
PineiroOrioli2015a,
Chantesana2018a},
see 
\cite{Gasenzer2011a,
Nowak2011a,
Nowak2012b,
Schole2012a,
Gasenzer2013a,
Ewerz2014a, 
Karl2017b,
Deng2018a,
Schmied:2018osf.PhysRevA.99.033611, 
Schmied2019a,
Groszek2020a.SciPostPhys.8.3.039,
Groszek2021a.PhysRevResearch.3.013212,
Spitz2021a.SciPostPhys11.3.060,
Siovitz:2023ius.PhysRevLett.131.183402, 
Huh:2023xso,
Siovitz2025a.PRA112.023304,
Noel2024:PhysRevD.109.056011,
Noel2025a.PRR7.033220,
Rasch:2025kna} 
in the context of vortex dynamics.

NTFPs have been proposed, in analogy to equilibrium critical phenomena, as a classification scheme for universal scaling dynamics far from equilibrium, see
\cite{Berges2015a,
Berges:2020fwq,
Mikheev2023a.EPJST232.3393,
Siovitz2023b}
for overviews.
Generalizing upon stationary dynamical processes such as fully developed turbulence, for which statistical observables like energy or particle-number distributions remain constant in time within a certain range of scales, NTFPs include the possibility of quasi-stationary evolution characterized by self-similar scaling in space and time.
This makes them particularly relevant for decaying turbulence, where transport processes such as inverse cascades arise during relaxation rather than in a stationary state.
In this picture, the time evolution can be viewed as a renormalization-group (RG) flow, classifying universal space-time scaling based on the symmetries of an underlying field model, with the potential to also gain a systematic description of the possible evolutions towards and away from the fixed-point solution.
Such scaling evolution subject to conservation laws was analyzed in the context of Bose condensation in  \cite{Svistunov1991a,Kagan1992a,Kagan1994a,Kagan1995a,Semikoz1995a.PhysRevLett.74.3093,Semikoz1997a}, cf.~also \cite{Chantesana2018a}.
Recent experiments, largely in ultracold gases, have explored various aspects of universal space-time scaling 
\cite{Johnstone2019a,
Eigen2018a,
Prufer2018a,
Erne2018b,
Navon2018a.Science.366.382,
Glidden:2020qmu,
GarciaOrozco2021aNoArxiv,
Lannig:2023fzf,
Martirosyan:2023mml,
Gazo2025a,
MorenoArmijos2024a,
Martirosyan2024a}.

%==================================
\begin{figure*}[t]
	\centering
	\includegraphics{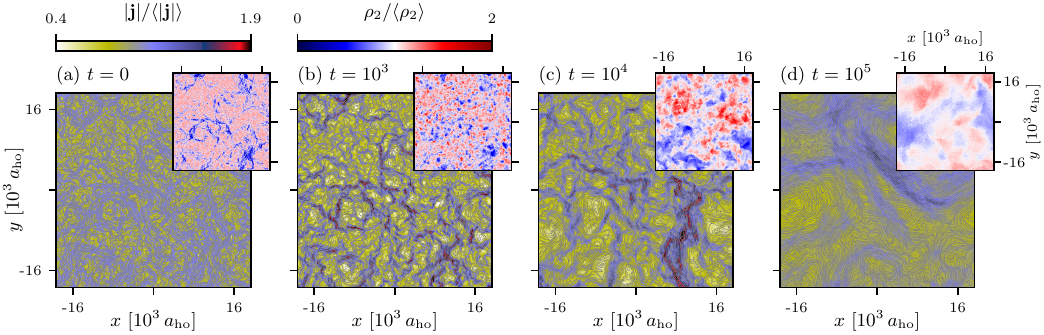}
	\caption{
		Normalized current $|\vb{j}|/\langle |\vb{j}| \rangle$ at four times $t\in[0,10^3,10^4,10^5](2\pi)/\omega_{z}$, with streamlines indicating the flow of $\vb{j}$.
		The system approaches space-time scaling near the anomalous NTFP within the time interval $t\approx(0.3\dots2)\cdot10^4(2\pi)/\omega_{z}$, corresponding to a decaying turbulent cascade, together with spatial scaling characteristics akin to 2d fully developed turbulence.
		The insets show the corresponding snapshots of the density normalized to its spatial average, $\rho_2/\langle \rho_2 \rangle$.
	}
	\label{fig:fig_1}
\end{figure*}
%==================================

%==============================================================================
\emph{Main result}.---%
We show that the anomalous NTFP known to exist in a dilute 2d superfluid \cite{Karl2017b,Rasch:2025kna} describes the decay of quantum turbulence, which exhibits long-range spatial scaling of the second moment of velocity circulation $\Gamma$ according to the classical Kraichnan-Kolmogorov power-law $\Gamma^2(r)\sim r^{8/3}$.
Moreover, we find the scaling of higher moments of the velocity circulation to show intermittent deviations consistent with the bifractal intermittency model studied in stationary IECs \cite{Zhu2023a.PhysRevLett.130.214001,Mueller2024.PhysRevLett.132.094002}.
In contrast, vortex separation is found to remain Gaussian distributed throughout the universal dynamics and does not exhibit intermittent behavior in time.
The NTFP we are focusing on in the closed system is characterized by vortex pattern coarsening and thus a growth of the inter-defect distance, $\ell_\text{v}\sim t^{\,\beta}$, with an anomalously small, subdiffusive exponent $\beta\approx1/5$, distinctly smaller than what is known from classical turbulence \footnote{%
In classical turbulence, this exponent is usually denoted as $\xi$, with $\xi/2=\beta$.}.
This process is associated with an inverse transport of the kinetic energy of incompressible turbulent fluid flow, exhibiting classical scaling, which decays and thereby expands further to lower momenta according to the conservation law associated with the NTFP.
Our results link superfluid turbulence in a closed 2d system, on the grounds of a high-resolution analysis in the spatial domain, to the concept of NTFPs in far-from-equilibrium quantum dynamics.

%==============================================================================
\emph{Non-thermal fixed points vs.~turbulence}.---%
For single-com\-ponent Bose gases, universal dynamics near NTFPs is typically dominated by nonlinear density excitations such as vortices, solitons, or kinks 
\cite{Gasenzer2011a, 
Nowak2011a,
Nowak2012b,
Schole2012a, 
Gasenzer2013a,
Ewerz2014a, 
Karl2017b,
Deng2018a, 
Johnstone2019a, 
Schmied:2018osf.PhysRevA.99.033611, 
Schmied2019a, 
Groszek2020a.SciPostPhys.8.3.039,
Groszek2021a.PhysRevResearch.3.013212,
Spitz2021a.SciPostPhys11.3.060,
Siovitz:2023ius.PhysRevLett.131.183402, 
Huh:2023xso,
Siovitz2025a.PRA112.023304,
Noel2024:PhysRevD.109.056011,
Noel2025a.PRR7.033220,
Rasch:2025kna}, with separate vortices being topologically protected.
The spatio-temporal coarsening of the field pattern is reflected in the loss of these excitations, and a length scale $\ell(t)\sim t^{\,\beta}$ is typically associated with the mean inter-defect distance.
In dilute, quasi-2d Bose gases, two distinct scaling exponents, $\beta\approx1/5$ and $\beta\approx1/2$ are observed, corresponding to an anomalous \footnote{The term \emph{anomalous} follows the RG convention, indicating that the scaling exponent deviates from its canonical value predicted by dimensional analysis.}
and a Gaussian NTFP, respectively
\cite{Schole2012a, 
Ewerz2014a, 
Karl2017b,
Reeves2017a.PhysRevLett.119.184502,
Deng2018a,
Johnstone2019a,
Groszek2020a.SciPostPhys.8.3.039,
Groszek2021a.PhysRevResearch.3.013212,
Spitz2021a.SciPostPhys11.3.060,
Noel2024:PhysRevD.109.056011,
Noel2025a.PRR7.033220,
Rasch:2025kna,
NoteTempQuenches}.
These reflect different vortex annihilation processes: $\beta=1/2$ arises from friction due to interaction with sound excitations, causing direct vortex-antivortex annihilations according to a two-body scattering rate.
In contrast, $\beta=1/5$ results for annihilations relying on three-vortex processes bringing the vortex pairs of opposite circulation sufficiently close to each other.
At early times and low temperatures, before vortex annihilations generate thermalizing sound noise, the sub-diffusive exponent dominates
\cite{Karl2017b,
Deng2018a,
Johnstone2019a,
Spitz2021a.SciPostPhys11.3.060,
Noel2024:PhysRevD.109.056011,
Noel2025a.PRR7.033220,
Rasch:2025kna}.

In turbulence, an important observable besides energy spectra are velocity increments, $\delta \vb{v}_{\vb{r}}(\vb{x}) = \vb{v}(\vb{x}+\vb{r})-\vb{v}(\vb{x})$, which follow distributions that are Gaussian for large $r=|\mathbf{r}|$ but develop heavy tails at smaller scales for fully developed 3d turbulence \cite{Frisch1995a}.
These deviations, termed intermittency \cite{Benzi1984a,Note3} are quantified by higher-order structure functions $S_{\!p}(r) = \langle |\delta\vb{v}_{\vb{r}}(\vb{x})|^p \rangle$, which scale as $S_{\!p}(r)\sim r^{\zeta_p}$ with $\zeta_1 = 1/3$ in a fully developed 2d IEC.
The higher exponents have been found both numerically \cite{Boffetta2000a} and experimentally \cite{Paret1998a} to scale as $\zeta_p = p/3$, implying the absence of intermittency, unlike 3d turbulence \cite{Belin1996a.PhysicaD93.52}.
Instead of velocity increments, however, the moments of the velocity circulation, 
\begin{align}
	\Gamma(r, \vb{x}) = \oint_{\mathcal{C}_{r,\vb{x}}} \vb{v} \dd{\vb{l}} \,,
	\label{eq:Circulation}
\end{align}
evaluated along a path $\mathcal{C}_{r,\vb{x}}$, e.g., a circle of radius $r$ around the point $\vb{x}$ \cite{Migdal1994a}, exhibit intermittency in the IEC, as seen in experiments \cite{Zhu2023a.PhysRevLett.130.214001} and simulations \cite{Mueller2024.PhysRevLett.132.094002}.
Velocity circulations are of particular interest in quantum turbulence since, by Stokes' theorem, they equal area integrals of quantized vorticity, i.e., the net circulation of vortices within a given area.

%==================================
\begin{figure*}[t]
	\centering
	\includegraphics{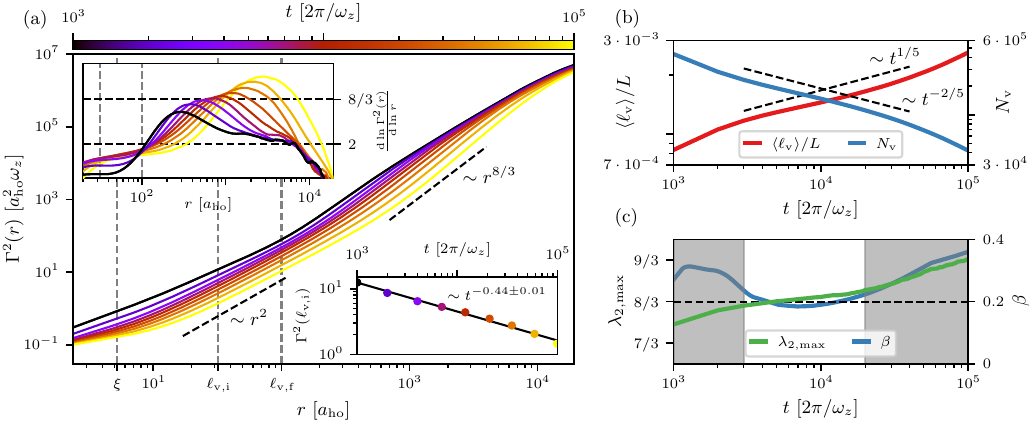}
	\caption{
		(a) Second moment $\Gamma^2(r)$ of the velocity circulation \eq{Circulation} about squares of side length $r$ at logarithmically spaced times in the interval $t\in\{10^3\dots10^{5}\}(2\pi)/\omega_z$.
		The dashed vertical lines indicate the healing length $\xi$ and the inter-defect distances $\ell_\mathrm{v,i}\approx32\,\aho$ and $\ell_\mathrm{v,f}\approx100\,\aho$ at the earliest and latest times in the interval, respectively.
		Below the scale set by $\ell_\text{v}(t)$, we recover an approximate scaling with $r^2$, which increases to Kraichnan-Kolmogorov scaling $\sim r^{8/3}$ on larger scales.
		The upper-left inset shows the local logarithmic derivative of $\Gamma^{2}$ calculated via finite differences over two grid-spacings.
		It also displays the transition from an approximate $r^2$ towards and $r^{8/3}$ power-law.
		The lower-right inset shows the decay of the second moment of circulation at $\ell_\mathrm{v,i}$ over time with exponent $\alpha=0.44(1)$.
		(b) Average inter-defect distance (red) and vortex number (blue), derived separately from the numerical data.
		At intermediate times, both follow power-laws with $\beta\approx1/5$ and $\alpha\approx2\beta$, respectively.
		(c) Comparison of the fitted time-dependent $\beta$ exponent extracted from the vortex number in (b) and the maximal local slope $\lambda_{2,\text{max}}$ from (a).
		The system approaches universal dynamics, indicated by $\beta\approx1/5$, at the highlighted intermediate times, which is accompanied by the buildup of a transient 2d turbulent Kraichnan-Kolmogorov scaling corresponding to a decaying IEC.
	}
	\label{fig:fig_2}
\end{figure*}
%==================================

%==============================================================================
\emph{2d Bose gas}.---%
We consider a single-component 2d dilute superfluid in the vicinity of the anomalous NTFP, exhibiting subdiffusive scaling ($\beta\approx1/5\ll1/2$) \cite{Karl2017b,Rasch:2025kna}.
Although a strictly 2d Bose gas cannot condense, it exhibits quasi-condensation characterized by quasi-long-range order up to the inter-defect scale $k_{\ell_\mathrm{v}}$.
Its dynamics is governed by the 2d Gross-Pitaevskii equation (GPE),
\begin{align}
	\label{eq:quasi2dGPE_dimless}
	\i \partial_{t}{\psi(\vb{r},t)} &= \left( - \nabla^2/2 + g_{2} \,\rho_2(\vb{r},t) \right) \psi(\vb{r},t) 
	\,,
\end{align}
where $\rho_2(\vb{r},t)=|\psi(\vb{r},t)|^2$ is the local 2d density and $g_{2}$ the dimensionless coupling.
We initialize the 2d Bose gas in a square volume with periodic boundary conditions, using an ensemble of $N_\mathrm{v}=1.4\cdot10^6$ randomly distributed (anti)vortices with zero net winding.
\Fig{fig_1}a shows an example of the corresponding normalized current, $\vb{j} = \rho_2 \vb{v} = \rho_2 \nabla \theta$, and density from $\psi=\sqrt{\rho_{2}}\e^{\i\theta}$.
The field is propagated by \eq{quasi2dGPE_dimless} in semi-classical Truncated-Wigner approximation (TWA) \cite{Blakie2008a}.
See Sect.~I of the Supplemental Material~\cite{SM} for details of parameters, units, initialization, and propagation.

%==================================
\begin{figure*}[t]
	\centering
	\includegraphics{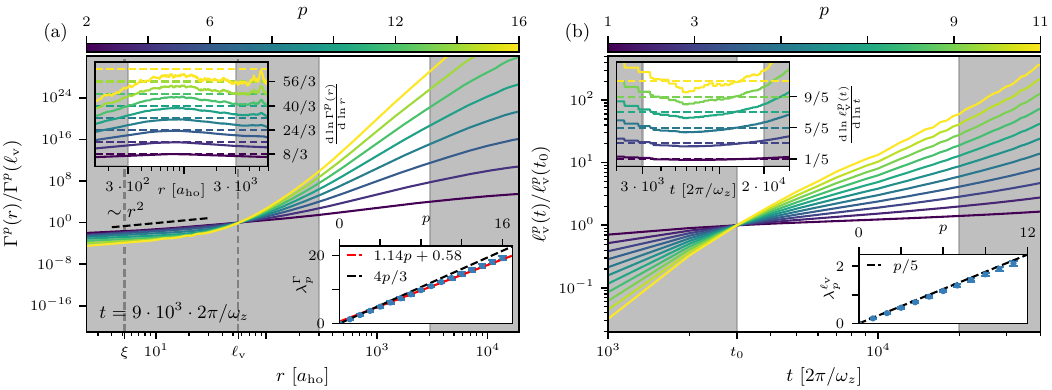}
	\caption{
		(a) $p$th-order moments of the velocity circulation, normalized at the inter-defect distance $\ell_\mathrm{v}$, at $t=9\cdot10^3(2\pi)/\omega_z$.
		Between the healing-length $\xi$ and the inter-defect scale $\ell_\mathrm{v}$ we recover scaling as $r^2$.
		The inertial range is set to the highlighted interval $300<r/\aho<3000$.
		The upper-left inset shows the logarithmic derivatives of the moments using finite difference over two grid-spacings, which approach power-laws within the inertial range, approximately given by $4p/3$ (dashed lines).
		The lower-right inset shows the averaged scaling exponents within the inertial range vs.~$p$, which exhibit intermittent deviations from $4p/3$ for higher moments.
		The deviations closely follow the linear function $1.14p+0.58$, as predicted for stationary IECs \cite{Zhu2023a.PhysRevLett.130.214001,Mueller2024.PhysRevLett.132.094002}.
		(b) Higher-order moments of the average inter-defect distance distribution as a function of time, normalized at $t_0=3\cdot10^3(2\pi)/\omega_z$.
		The upper-left inset displays the logarithmic derivatives, obtained similarly to \(\beta\) in \Fig{fig_2}c, which approach power-law behavior within the inertial range.
		In the lower-right inset the extracted scaling exponents closely follow $p/5$, indicating that the inter-defect distribution is nearly Gaussian and does not show `intermittent' behavior.
	}
	\label{fig:fig_3}
\end{figure*}
%==================================

%==============================================================================
\emph{Vortex pattern coarsening close to the anomalous NTFP}.---%
Fig.~\fig{fig_1} shows four snapshots of the system for an exemplary run, see \cite{VideosNTFP2dTurb} for a video.
The current exhibits vortex clusters surrounded by strong coherent flows, which weaken in time along with the coarsening of the vortex pattern, also reflected in the growing density variation patterns.
Choosing a high initial vortex number, transport to lower wave numbers and thus the mutual annihilation of vortices and antivortices sets in quickly, while achieving a sufficient large-$r$ resolution of the circulation correlators.
By the time $t\approx10^{3}(2\pi)/\omega_{z}$, shown in \Fig{fig_1}b, the dynamics has lead from the non-equilibrium initial condition to the build-up of a turbulent cascade, which exhibits decay due to the lack of external driving.

We find the coarsening to be characterized by an algebraic time evolution of the mean inter-defect, i.e., the shortest distance between a vortex and its neighbor, $\ell_\mathrm{v}(t) \sim t^{\,\beta}$, and of the mean vortex number, $N_\mathrm{v}(t)\sim t^{-2\beta}$, cf.~\Fig{fig_2}b.
After an initial buildup phase, the scaling exponent takes the subdiffusion-type value $\beta\approx1/5$, consistent with the anomalous NTFP described in \cite{Karl2017b,Rasch:2025kna} and remains approximately constant within the time interval $3\cdot10^3\lesssim (2\pi)^{-1}\omega_{z}t\lesssim2\cdot10^4$.
At later times, cf.~\Fig{fig_1}d, the vortex number decay departs from anomalous scaling due to an increased presence of sound excitations.
These excitations reflect the approximate effective heating induced by vortex annihilations in the closed system, raising compressible excitations in the region of large wave numbers.
The growing compressible component fosters two-body vortex annihilations, with $\beta$ effectively increasing (cf.~\Fig{fig_2}b) until the system eventually reaches diffusion-type scaling with $\beta\approx1/2$ \cite{Karl2017b,Rasch:2025kna}.
The time-local scaling exponent $\beta(t)$, obtained by fitting a power law to the decaying vortex number over half an order of magnitude symmetrically around $t$, is shown in \Fig{fig_2}c (blue line).
Unlike the anomalous scaling observed over up to two orders of magnitude in time in \cite{Rasch:2025kna}, the higher initial vortex density here accelerates pair annihilations, whose sound emissions trigger an earlier crossover from $\beta\approx1/5$ to diffusion-type scaling.

%==============================================================================
\emph{Decaying inverse cascade}.---%
Spatio-temporal scaling close to an NTFP is typically studied via correlation functions such as the angle-averaged single-particle momentum spectrum, $n(k,t)$, which near the fixed point takes the scaling form $n(k,t)= [t/t_{0}]^{\alpha}f_\text{s}([t/t_{0}]^{\,\beta}k)$, with universal scaling function $f_\text{s}(k)=n(k,t_{0})$ \cite{PineiroOrioli2015a}.
The temporal scaling dimensions $\beta$ of $k$ and $\alpha$ of $f_\text{s}$ are usually constrained by a conservation law; e.g., of total particle number, which implies $\alpha=d\beta$ in $d$ dimensions, while $\beta$ is fixed by the fixed point's universality class.
Positive $\beta>0$ corresponds to transport toward lower momenta, with a coarsening length scale $\ell$ parametrizing the scaling function, e.g., $f_\text{s}(k)\sim (k_{\ell}^{2}+k^{2})^{-\kappa/2}$, $k_{\ell}\sim\ell^{-1}$ \cite{Karl2017b}.
For such $f_\text{s}$, the high-momentum tail decays algebraically in time, $n(k\gg k_{\ell},t)\sim [t/t_{0}]^{\alpha-\beta\kappa}k^{-\kappa}$, if $\kappa>\alpha/\beta$ \cite{Chantesana2018a}. 
Fig.~1 of the Supplemental Material \cite{SM} shows, the single-particle spectrum of our vortex gas obeys the above scaling form, with a subdiffusive exponent $\beta=0.22(3)$, and the relation $\alpha\approx 2 \beta$ signals a particle-number conserving inverse flow towards lower momenta in the form of a decaying cascade.
At small momenta, while its extent into the infrared (IR) grows, higher occupancies are built up as $n(k\ll k_{\ell,0}\tau^{-\beta},t)\sim\tau^{\alpha}$, with IR characteristic scale $k_{\ell,0}=1/\ell_\text{v}(0)\approx0.05/\aho$.

To better quantify the decaying cascade, we analyze the flow pattern in position space, where IR statistics are supported by more grid points than in momentum space.
Specifically, we study moments of the velocity circulation \eq{Circulation}, for which we integrate the flow around square contours $\mathcal{C}_{r,\vb{x}}$ of edge length $r$, with the lower-left corner at $\vb{x}$, to match the grid symmetry.
Higher-order moments $\Gamma^p(r) \equiv \langle \mathcal{N}^{-1}\sum_\mathbf{x}|\Gamma(r, \vb{x})|^p \rangle$ are evaluated by averaging over all contours anchored at $\mathcal{N}=16384^2$ lattice points and $20$ TWA runs.
For fully developed, non-intermittent turbulence, the moments of the velocity circulation exhibit power-law scaling $\Gamma^p(r) \sim r^{\lambda_p}$ within the inertial range.
Kolmogorov's theory of 3d turbulence predicts the exponents to scale as $\lambda_p=4p/3$, which also holds for an IEC in 2d, as can be inferred from the energy spectrum $E(k) \sim k^{-5/3}$ in Kraichnan-Leith-Batchelor theory \cite{Kraichnan1967a,Leith1968a.PhysFl11.671,Batchelor1969a.PhysFl12.II-233}.

In \Fig{fig_2}a, the second moment $\Gamma^2(r)$ of the velocity circulation is shown over two orders of magnitude in time.
At length scales $r\lesssim\ell_\mathrm{v,f}=\ell_\text{v}(t_\text{f})$, it scales as $\Gamma^{2}(r)\sim r^2$, cf.~the local logarithmic derivative in the upper-left inset.
The systematic deviation from the expected value is likely due to the rather high vortex density.
Evaluated at the initial-time inter-defect distance $\ell_\mathrm{v,i}\equiv\ell_\text{v}(t=t_\text{i})\approx32\,\aho$, $t_\text{i}=10^{3}(2\pi)/\omega_{z}$, the circulation decays as $\Gamma^2(\ell_\mathrm{v,i})\sim t^{-0.44\pm0.01}$, cf.~lower-right inset, consistent with the decrease of the defect number $N_\text{v}(t)$ (\Fig{fig_2}b, blue line).
The vortex number scaling is recovered since the moments do not distinguish vortices from antivortices, and in a self-similarly coarsening vortex ensemble, the second moment remains approximately constant at the inter-defect distance scale, $\Gamma^{p}(\ell_\mathrm{v}(t))\approx \mathrm{const.}$.
Hence, with $\ell_\mathrm{v}(t)\sim t^{\,\beta}$, one expects $\Gamma^{2}(\ell_\mathrm{v,i})=\Gamma^{2}(\ell_\mathrm{v}(t)(t/t_\mathrm{i})^{-\beta})\sim t^{-2\beta}\sim N_\text{v}(t)$.
At length scales $r\gg\ell_\mathrm{v,f}=\ell_\text{v}(t_\text{f})$, a steeper scaling emerges, with exponent $\lambda_2 \approx 8/3$ at intermediate times, cf.~the upper-left inset.

Our results demonstrate a spatially varying exponent in this regime, yet it is striking that values around $\lambda_{2}=\mathrm{d}\ln\Gamma^{2}(r)/\mathrm{d}r\approx8/3$ appear to be attained by the maximum slope $\lambda_{2,\mathrm{max}}(t)$ during the time interval $\approx(3..20)\,t_\text{i}$ (white area in \Fig{fig_2}c), coinciding with the regime where $\beta\approx1/5$ signals the approach of the anomalous NTFP.
Hence, alongside the scaling exponent $\beta$, the emergence of universal dynamics near the anomalous NTFP is accompanied by the buildup of a decaying inverse cascade, which exhibits spatial scaling with $\lambda_2=8/3$.
This is also found by explicitly rescaling the circulation probability distribution function, cf. Fig.~4 of the Supplemental Material \cite{SM}.
There, in Sect.~IV\,B, further details on the transient behavior of \(\lambda_{2,\mathrm{max}}(t)\) at the single-run level are given.

%==============================================================================
\emph{Intermittency}.---%
By evaluating higher-order moments of $\Gamma$, we quantify intermittent deviations from $\lambda_{p}$ increasing linearly in $p$.
\Fig{fig_3}a shows $\Gamma^p(r)/\Gamma^p(\ell_\text{v})$ for even $2\leq p\leq16$ at $t=9\,t_\mathrm{i}$, when universal scaling dynamics is fully developed.
The probability distribution functions in Fig.~5 of the Supplemental Material \cite{SM} confirm statistical convergence for all chosen $p$.
Below the inter-defect scale, $r\ll\ell_\mathrm{v}$, the area-law scaling $\sim r^2$ is recovered for all moments, whereas within the highlighted inertial range $\{3\dots30\}\cdot10^{2}\,\aho$, steeper power-law scaling emerges.
In the upper-left inset, the local logarithmic derivatives of these moments approach values close to $\lambda_{p}=4p/3$ (dashed lines) within the inertial range, as expected from classical turbulence.
Fitting power-laws in intervals of width $240\,\aho$ across the inertial range and averaging, we extract scaling exponents which are shown, vs.~$p$, in the lower-right inset.
Compared to $4p/3$, higher moments display intermittent deviations, reflecting the spatially non-Gaussian vortex statistics in the inertial range.
The red dashed line shows the bifractal intermittency model, $\lambda_{p} = h p + (2-D)$ for $p>3$, using the experimentally determined H\"older exponent $h=1.14(2)$ and fractal dimension $D=1.42$ \cite{Zhu2023a.PhysRevLett.130.214001}, which are consistent with our results.
In this model, the flow is assumed to consist of two distinct fractal subsets, each with its own scaling exponent, yielding a piecewise-linear dependence of structure function exponents on moment order \cite{Frisch1995a}.

We perform a similar analysis of the temporal scaling of the higher moments of the average inter-defect distribution $\ell_\mathrm{v}$ in \Fig{fig_3}b for all integer $1\leq p\leq11$, cf. Fig.~6 of the Supplemental Material \cite{SM} for statistical convergence.
Within the universal time interval, all higher moments show power-law scaling with exponents $p/5$, cf.~the upper-left inset, without clear intermittency deviations, cf.~the lower-right inset.
This indicates that the inter-defect distribution remains Gaussian within the universal interval up to the $p=11$ moment and no `temporal intermittency' is observed.
Hence, spatial intermittency on inertial-range scales does not affect the vortex distribution on inter-defect scales.
Whether this persists for $n$th-nearest neighbor distances, which eventually probe inertial-range scales, remains an open question.

%==============================================================================
%==============================================================================
\emph{Conclusions}.---%
Starting from a large, randomly sampled vortex ensemble with zero net circulation in a dilute 2d superfluid Bose gas, we have shown that the ensuing closed-system dynamics -- conserving both energy and particle number -- develops the spatial scaling characteristics of an inverse Kraichnan-Kolmogorov cascade.
Vortex annihilations lead to a progressive loss of defects and induce a self-similar transport of the single-particle momentum distribution toward lower wave numbers, thereby increasing the occupation of the lowest momenta and signalling the approach of universal dynamics close to an NTFP.
The evolution bears the characteristics of decaying 2d turbulence, in which the vortex pattern coarsens in time, with the mean inter-vortex distance growing algebraically as $\ell_\mathrm{v}(t)\sim t^{\,\beta}$.
During the same temporal window, the coarsening exponent approaches $\beta\approx1/5$, while the spatial scaling of the velocity circulation follows $\Gamma(r)\sim r^{4/3}$, consistent with decaying Kraichnan-type turbulence.
The observed coarsening exponent is significantly smaller than the value $\beta\approx0.35$ characteristic of classical 2d turbulence with dominant three-vortex scattering \cite{Carnevale1991a.PhysRevLett.66.2735,Bokhoven2007a.PhysFl19.046601,Sire2010a.PhysRevE.84.056317,Note1}.
Moreover, the $p$th-order moments of $\Gamma$ exhibit intermittent deviations from the scaling $\sim r^{4p/3}$, consistent with intermittency reported in classical, non-decaying IECs \cite{Zhu2023a.PhysRevLett.130.214001,Mueller2024.PhysRevLett.132.094002}.
In contrast, the temporal scaling of the average inter-defect distance was found to remain non-intermittent up to $p=11$.
In summary, our results demonstrate that decaying superfluid turbulence in a closed system can undergo universal coarsening dynamics while exhibiting key characteristics of classical turbulence.
They establish a connection between NTFPs and turbulence in quantum fluids, with dynamics governed by non-stationary, self-similar scaling in space and time.
This raises the question how analogous dynamical scaling attractors in classical fluids, beyond the quasi-stationary evolution of decaying turbulence, can be understood in terms of NTFPs, which would be of far-reaching interest.
In this context, clarifying the microscopic origin of the exponent $\beta\approx1/5$ of the quantum fluid and its difference to the respective exponents reported for classical turbulence is of central interest.
It has been associated with $3$-vortex processes in a kinetic description of their scattering \cite{Karl2017b}, and there exists an ansatz to connect it to scaling in an effective sine-Gordon-type theory \cite{Heinen2023a}.

%==============================================================================
%==============================================================================
\emph{Acknowledgments}.
The authors thank A.~Bradley, K.~Chandrashekara, L.~Chomaz, G.~Eyink, P.~Heinen, W.~Kirkby, H.~K{\"o}per, G.~Krstulovic, A.-M.~Oros, D.~Proment, I.~Siovitz, and G.~A.~Williams for discussions and collaboration on related topics.
The authors acknowledge support by the Deutsche Forschungsgemeinschaft (DFG, German Research Foundation), through SFB 1225 ISOQUANT (Project-ID 273811115), grant GA677/10-1, and under Germany's Excellence Strategy -- EXC 2181/1 -- 390900948 (the Heidelberg STRUCTURES Excellence Cluster); by the state of Baden-W{\"u}rttemberg through bwHPC, the data storage service SDS{@}hd supported by the Ministry of Science, Research and Arts Baden-W{\"u}rttemberg (MWK), and the DFG through grants INST 35/1503-1 FUGG, INST 35/1597-1 FUGG, and INST 40/575-1 FUGG (SDS, Helix, and JUSTUS 2).
We thank the Institute for Nuclear Theory at the University of Washington for its kind hospitality and stimulating research environment. This research was supported in part by the INT's U.S. Department of Energy grant No. DE-FG02- 00ER41132.

%==============================================================================

\begin{appendix}

\vspace{1cm}
\begin{center}
\textbf{APPENDIX}
\end{center}
\setcounter{equation}{0}
\setcounter{table}{0}
\makeatletter

%==============================================================================
%==============================================================================
\section{Single-component dilute superfluid Bose gas}
\label{app:BoseGas}
%
%==============================================================================
\subsection{Gross-Pitaevskii model in a quasi-2d geometry}
\label{app:2dGPEmodel}
The temporal evolution of a 3d single-component Bose gas at zero temperature is given by the nonlinear Gross-Pitaevskii equation (GPE),
\begin{align}
	\label{eq:GPE}
	\i \hbar \pdv{\Psi(\vb{r},t)}{t} &= \left( - \frac{\hbar^2}{2m} \nabla^2 + V_\mathrm{ext}(\vb{r}) + g \rho(\vb{r},t) \right) \Psi(\vb{r},t) \,,
\end{align}
with $\rho(\vb{r},t) = |\Psi(\vb{r},t)|^2$ the local 3d particle density with mass $m$ and $V_\mathrm{ext}(\vb{r})$ the trapping potential.
At low energies, the microscopic short-range interaction in 3d can be approximated by a local contact interaction with coupling $g=4\pi\hbar^2a_\mathrm{s}/m$, in terms of the 3d $s$-wave scattering length $a_\mathrm{s}$.
For a homogeneous density $\rho(\vb{r})=\rho$, the chemical potential in mean-field approximation is given by $\mu=g\rho$, from which the healing length $\xih=\hbar/\sqrt{2m\mu}$ is obtained.

To study quantum vortices in a quasi-2d geometry, we introduce an external trapping potential $V_\mathrm{ext}(\vb{r})=m\omega_z^2 z^2 /2$ along the $z$-direction.
This confines the system, using a trapping frequency $\omega_z$, to the characteristic width of the harmonic oscillator $\aho=\sqrt{\hbar/(m\omega_z)}$.
If $\aho\lesssim\xih$, the dynamics along the $z$-direction is effectively frozen out and the system is, to a good approximation, in the single-particle ground state of the harmonic oscillator along the $z$-direction,
\begin{align}
	\label{eq:harmonic_gs}
	h(z) = \left(\pi\aho\right)^{-1/4} \exp{-\frac{z^2}{2\aho^2}} \,.
\end{align}
Factorizing the wave function $\Psi(\vb{r},t)=\psi(\vb{r}_\perp,t) h(z)$ into the planar contribution and $h(z)$, the $z$-dependence can be integrated out explicitly.
This yields the quasi-2d GPE,
\begin{align}
	\label{eq:quasi2dGPE}
	\i \hbar \pdv{\psi(\vb{r}_\perp,t)}{t} &= \left( - \frac{\hbar^2}{2m} \nabla_\perp^2 + g_2 \rho_2(\vb{r}_\perp,t) \right) \psi(\vb{r}_\perp,t) \,,
\end{align}
with planar density $\rho_2(\vb{r}_\perp,t)\equiv|\psi(\vb{r}_\perp,t)|^2$ and renormalized contact interaction $g_2 = g/(\sqrt{2\pi} \aho)=\sqrt{8\pi}\hbar^{2} a_\mathrm{s}/(m\aho)$.
For a homogeneous density $\rho_2(\vb{r}_\perp,t)=\rho_2$, the chemical potential is given by $\mu_{2}=g_{2}\rho_{2}=\sqrt{8\pi}\hbar^{2} a_\mathrm{s} \rho_2/m$, which defines the 2d healing length $\xih_{2}=1/k_{\xih_{2}}=\hbar(2m\mu_{2})^{-1/2}=(4\sqrt{2\pi}a_\mathrm{s}\rho_{2}/\aho)^{-1/2}$.
The latter sets the scale on which density modulations are restored in the system and thus measures the vortex core size.

We solve the dynamics of the system using the semiclassical Truncated-Wigner approximation (TWA), in which initial field configurations with randomly sampled added noise are propagated with the above 2d GPE as classical equation of motion.
The random noise is added to the phase of the Bogoliubov quasiparticle mode excitations with dispersion $\omega(\mathrm{k})$, 
\begin{align}
	\label{eq:BogDispersion}
	\omega(\vb{k})^2 = \epsilon_k \left( \epsilon_k + 2 \mu_2 \right) \,,
\end{align}
with single-particle energy $\epsilon_k=\hbar^2 k^2/(2m)$.

%==============================================================================
\subsection{Units and parameters}
\label{app:params}
Eq.~(2) in the main text is expressed in dimensionless form, with length in units of $[L]=\aho$, time in $[T]=1/\omega_z$, and energy in $[E]=\hbar\omega_z$, with harmonic oscillator length $\aho$ and trapping frequency $\omega_z$ of the quasi-2d confinement along the $z$-axis.
This is achieved by defining dimensionless coordinates, couplings and fields:
\begin{align}
	\tilde{x} = \frac{x}{\aho} \,, \qquad
	\tilde{k} = k \aho \,, \qquad
	\tilde{t} = \omega_z t \,, \notag \\
	\tilde{\psi}(\tilde{\vb{r}}_\perp,\tilde{t}) 
	= \psi(\vb{r}_\perp,t) \aho \,, \qquad
	\tilde{a}_\mathrm{s} 
	= \frac{a_\mathrm{s}}{\aho} 
	\,.
\end{align}
This gives the dimensionless 2d coupling strength $\tilde{g}_2 = g_2 m /\hbar^2 =\sqrt{8\pi}\tilde{a}_\text{s}$.
Inserting this into \eqref{eq:quasi2dGPE} yields the dimensionless quasi-2d GPE, 
\begin{align}
	\i \pdv{\tilde\psi(\tilde{\vb{r}}_{\perp},\tilde{t})}{\tilde{t}} 
	&= \left( - \frac{1}{2} \tilde\nabla^2 
	+ \sqrt{8\pi} \tilde{a}_\mathrm{s} \tilde\rho_2(\tilde{\vb{r}}_{\perp},\tilde{t}) \right) \tilde\psi(\tilde{\vb{r}}_{\perp},\tilde{t}) \,.
	\label{eq:app:quasi2dGPE_dimless}
\end{align}
and after dropping the tildes and $\perp$ Eq.~(2) of the main text.

In the following we list the parameter values chosen for the propagation of \eqref{eq:app:quasi2dGPE_dimless}.
In order to observe the anomalous NTFP we adopted the \emph{ultradilute} parameter regime introduced in \cite{Karl2017b,Rasch:2025kna}.
We start from the particle number $N=\rho_2 L^2 = 3.2\cdot10^9$ and the dimensionless 2d coupling strength $\tilde{g}_2 = \sqrt{8\pi}\tilde{a}_\text{s}=1.5\cdot10^{-5}$ as chosen, for $m=1/2$, in \cite{Karl2017b}.
Together with the linear system size $\tilde{L} = L/\aho \approx 1612$ chosen in \cite{Rasch:2025kna}, the particle density is $\tilde\rho_2=N/\tilde{L}^2 \approx 1232$, which results in a dimensionless chemical potential
\begin{align}
	\tilde{\mu}_2 = \tilde{g}_2 \tilde{\rho}_2 \approx 0.0184 \,.
\end{align}
This choice satisfies the validity condition of the quasi-2d approximation, $\tilde{\xih}_{2} = 1/\sqrt{2\tilde{\mu}_2} \approx5.2$, to be sufficiently larger than $1$.
The 3d diluteness parameter at the transverse peak density is
\begin{align}
	\eta = \sqrt{\tilde\rho_3 \tilde{a}_\mathrm{s}^3} 
	= \frac{\sqrt{\tilde\rho_2 \tilde{g}_2^3}}{2^{1/4}4\pi} \approx 1.4 \cdot 10^{-7} \,,
\end{align}
which quantifies the high diluteness of the system achieved by weak interactions and large occupancies.
Since we perform our simulations on $16384^2$ numerical lattices, we enlarged the system to $\tilde{L}\approx3.87\cdot10^4$ and $N\approx1.84\cdot10^{12}$, keeping the density and the chemical potential fixed.

%==============================================================================
\subsection{Initialization and propagation}
\label{app:TWA}
We initialize the 2d Bose gas in a square volume $\tilde{V}=\tilde{L}^{2}$ with periodic boundary conditions, on a $\mathcal{N}=16384^2$ numerical lattice, with an ensemble of $N_\mathrm{v}=1.4\cdot10^6$ randomly distributed vortices and antivortices with net zero winding number.
We choose the high initial vortex number to initiate the buildup of a spatial turbulent cascade and achieve a sufficient large-$r$ resolution of the circulation correlators, cf.~\App{DefectNumber} for a discussion of the further optimization of the defect number.

The field configuration is propagated using \Eq{app:quasi2dGPE_dimless} within the semi-classical TWA \cite{Blakie2008a}. 
For this, we prepare the system by adding noise to the initial Bose field, in the form of an amplitude $\sqrt{1/2}$ and a random phase in each of the homogeneous-density Bogoliubov quasiparticle momentum modes with energy \eqref{eq:BogDispersion}.
The real-time evolution is computed up to $t_\mathrm{f}=10^5(2\pi)/\omega_z$, with time step $\dd{t}=0.1(2\pi)/\omega_z$, using a split-step Fourier algorithm which accounts for the periodic boundary conditions.
The resulting correlation functions are evaluated by averaging over $20$ runs performed in a parallelized fashion using GPU clusters.
We emphasize that, other than in \cite{Mueller2024.PhysRevLett.132.094002}, we do not suppress compressible (sound) excitations, neither in the initial state nor during the runs.

%==================================
\begin{figure*}[t]
	\centering
	\includegraphics{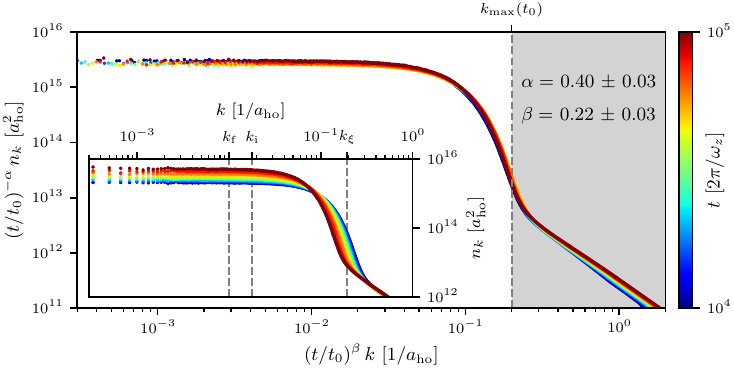}
	\caption{
		We show ten occupation number spectra \(n_k\) logarithmically spaced in time in the interval \(t\in[10^4,10^5](2\pi)/\omega_z\) in the lower-left inset.
		In the main plot, the spectra are rescaled using the scaling exponents \(\alpha=0.40\pm0.03\) and \(\beta=0.22\pm0.03\) demonstrating self-similar evolution in time.
		The UV cutoff at the reference time \(t_0=10^4(2\pi)/\omega_z\) is set to \(k_\mathrm{max}(t_0)=2\cdot10^{-1}/\aho\).
	}
	\label{fig:fig_8}
\end{figure*}
%==================================

%==============================================================================
%==============================================================================
\section{Self-similar evolution of the occupation number spectra close to the non-thermal fixed point}
\label{app:SelfSimilarity}
In the main text, we have demonstrated the emergence of universal dynamics by the presence of coarsening manifest in the anomalously slow power-law decay of the mean vortex number and, equivalently, the algebraic growth of the mean inter-defect distance as $\ell_\text{v}(t)\sim t^{\,\beta}$, with an exponent \(\beta=1/5\), as introduced in \cite{Karl2017b}, seen experimentally in \cite{Johnstone2019a}, and extended to dipolar systems in \cite{Rasch:2025kna}.
Beyond this, universal space-time scaling can be observed in the angle-averaged single-particle spectrum $n(k,t)$,
\begin{align}
	\label{eq:AngleAveragedOccNumber}
	n(k,t) &= \int \dd{\Omega} \langle \psi^*(\vb{k}-\langle\vb{k}\rangle,t) \,\psi(\vb{k}-\langle\vb{k}\rangle,t) \rangle 
	\,,
\end{align}
with $k=|\vb{k}|$, angular measure \(\dd{\Omega}=\dd\cos\theta_\mathbf{K}\dd\varphi_\mathbf{K}/(4\pi)\), $\mathbf{K}=\vb{k}-\langle\vb{k}\rangle$, and \(\langle\vb{k}\rangle\) the mean total momentum that is randomly imprinted when defects are sampled.
We emphasize that performing the angular integration about the centre-of-momentum mode $\langle\vb{k}\rangle$ turned out to be crucial in order to recover self-similar scaling \cite{Rasch:2025kna}.

In the vicinity of a NTFP, the single-particle spectrum takes the universal scaling form
\begin{align}
	\label{eq:ScalingHypothesis}
	n(k,t) = (t/t_0)^\alpha n([t/t_0]^\beta k,t_0) \equiv (t/t_0)^\alpha f_\mathrm{s}(k/k_\ell(t)) 
	\,,
\end{align}
with universal scaling function $f_\mathrm{s}(k)=n(k,t_{0})$ defined at the reference time $t_0$ and scaling exponents $\alpha$ and $\beta$ characterizing the fixed point.
The IR momentum scale decreases as a power-law $k_\ell(t) \sim (t/t_0)^{-\beta}$, corresponding to spatial coarsening, i.e., the growth of the characteristic length scale
\begin{align}
	\ell(t) \sim k_\ell(t)^{-1} \sim (t/t_0)^\beta \,.
\end{align}
\Fig{fig_8} shows the rescaling of ten spectra logarithmically spaced in time (lower-left inset) over the interval \(t\in[10^4,10^5](2\pi)/\omega_z\).
They are rescaled in the main plot using the scaling exponents \(\alpha=0.40\pm0.03\) and \(\beta=0.22\pm0.03\), obtained with a least-square routine put forward in \cite{PineiroOrioli2015a}.
These exponents agree with the subdiffusive scaling expected in the vicinity of the anomalous NTFP described in \cite{Karl2017b,Rasch:2025kna}.
The scaling form \eq{ScalingHypothesis} holds below the UV cutoff, \(k\lesssim k_\mathrm{max}(t_0)=2\cdot10^{-1}/\aho\), at the reference time \(t_0\).
Self-similar scaling, however, is seen to be most pronounced in the time interval, \(t\in[10^4,10^5](2\pi)/\omega_z\) cf.~\Fig{fig_8}, which exceeds the universal interval extracted from coarsening, \(t\in[3\cdot10^3,2\cdot10^4](2\pi)/\omega_z\) cf.~Fig.~2 of the main text.
This difference is attributed to the scaling function not being fully developed when coarsening in the vortex number is already observed.

%==============================================================================
%==============================================================================
\section{Incompressible energy spectrum}
\label{app:IncompressibleEnergy}
%

%==================================
\begin{figure*}[t]
	\centering
	\includegraphics{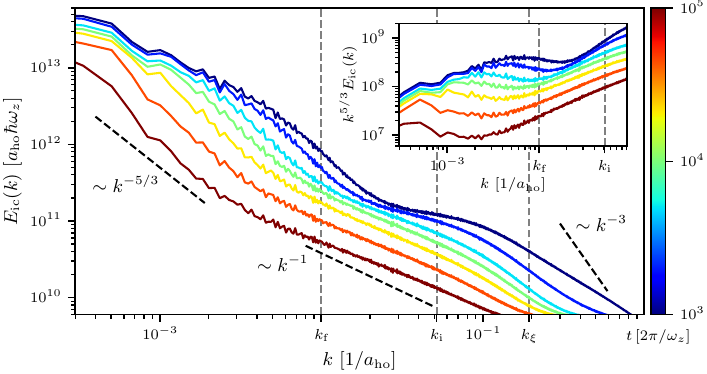}
	\caption{
		Incompressible energy spectra $E_\mathrm{ic}(k)$ at logarithmically spaced times in the interval $t\in[10^3,10^5](2\pi)/\omega_z$ with the healing scale $k_\xi$ and the initial $k_\mathrm{i}$ and final $k_\mathrm{f}$ inter-defect scales marked by vertical gray-dashed lines.
		For comparison we further show the expected scaling of Kraichnan-Kolmogorov turbulence $\sim k^{-5/3}$, of the vortex' velocity field $\sim k^{-1}$, and of the vortex core $\sim k^{-3}$.
	}
	\label{fig:fig_14}
\end{figure*}
%==================================

%==================================
\begin{figure*}[t]
	\centering
	\includegraphics[width=\textwidth]{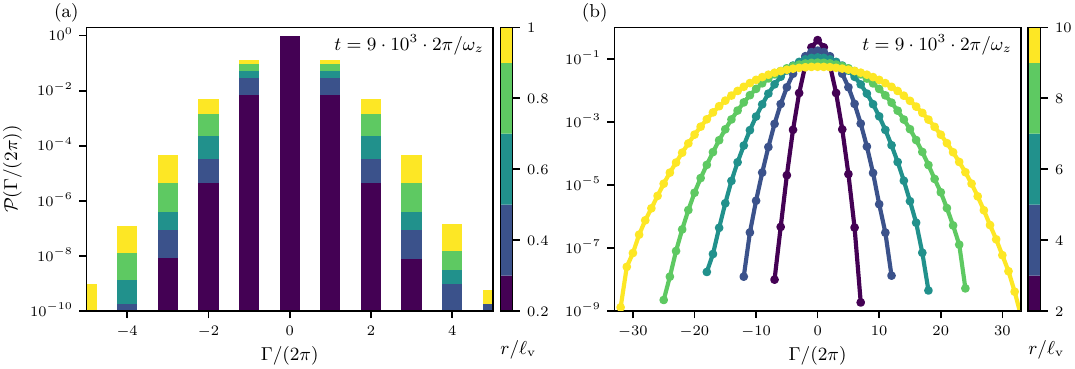}
	\caption{
		Probability density distributions of the mean circulation $\Gamma_{r}=\Gamma(r)$ around squares of side length $r$ in units of $2\pi$, at time $t=9\cdot10^{3}(2\pi)/\omega_{z}$, for five values of $r/\ell_\text{v}$ chosen as (a) $\in\{0.2,0.4,0.6,0.8,1.0\}$ and (b) $\in\{2,4,6,8,10\}$, respectively.
	}
	\label{fig:fig_11}
\end{figure*}
%==================================

Throughout this work we have discussed the emergence of Kraichnan-Kolmogorov scaling in the moments of the velocity circulation, cf.~Fig.~2 of the main text, since performing the analysis in position space provides a higher resolution of IR scales.
Notwithstanding this, turbulent scaling is usually also studied using the radial spectral energy distribution, which, in a stationary inverse cascade of a classical incompressible fluid in $d=2$, exhibits the scaling law $E(k)\sim k^{-5/3}$ \cite{Kraichnan1967a,Leith1968a.PhysFl11.671,Batchelor1969a.PhysFl12.II-233}.
For our dilute superfluid, we expect to recover the scaling behavior in the incompressible energy spectrum which we can determine by decomposing the kinetic energy into a `classical', $E_\mathrm{cl}$, and a `quantum' part, $E_\mathrm{q}$, by means of the Madelung transformation of the field in terms of density and phase,
\begin{align}
	E_\mathrm{kin} &= \frac{1}{2} \int \dd[2]{r} \abs{\nabla \psi(\vb{r})}^2 \notag \\
  &= \underbrace{\frac{1}{2} \int \dd[2]{r} \abs{\sqrt{\rho_2(\vb{r})} \, \vb{v}(\vb{r})}^2}_{E_\mathrm{cl}} + \underbrace{\frac{1}{2} \int \dd[2]{r} \abs{\nabla \sqrt{\rho_2(\vb{r})}}^2}_{E_\mathrm{q}} \,.
\end{align}
The `quantum' part $E_\mathrm{q}$ gives rise to the `quantum pressure' in the hydrodynamic description of the Bose gas and reflects the system's response to spatial gradients in the density \cite{Nore1997a.PhysRevLett.78.3896}.
One furthermore Helmholtz decomposes $\vb{w}_\mathrm{cl}=\sqrt{\rho_2} \, \vb{v} = \vb{w}_\mathrm{ic} + \vb{w}_\mathrm{c}$ into a divergence-free, i.e., incompressible, and a curl free, compressible contribution, whereby the amplitude factor guarantees that the respective kinetic energies add up to $E_\mathrm{cl}$ \cite{Nore1997a.PhysRevLett.78.3896}.
The contribution to $E_\mathrm{cl}$ from the divergence free flow yields the incompressible kinetic energy
\begin{align}
	E_\mathrm{ic} &= \frac{1}{2} \int \dd[2]{r} \abs{\vb{w}_\mathrm{ic}(\vb{r})}^2 = \frac{1}{2\pi} \int \dd{k} E_\mathrm{ic}(k) \,,
\end{align}
as well as its corresponding energy spectrum $E_\mathrm{ic}(k)$, which is shown in \Fig{fig_14} at logarithmically spaced times.
Deep in the IR, i.e., below the inter-defect scale, the spectra exhibit a steeper scaling following the Kraichnan-Kolmogorov power-law $\sim k^{-5/3}$ as can also be taken from the spectra multiplied by $k^{5/3}$ shown in the upper-right inset.
At intermediate scales, between the inter-defect and the healing scale, we find the scaling $\sim k^{-1}$ of the velocity field of a single vortex.
Beyond the healing scale, one would expect the vortex core scaling $\sim k^{-3}$, which we, however, cannot reproduce due to our limited numerical resolution in the UV.

%==================================
\begin{figure*}[t]
	\centering
	\includegraphics[width=\textwidth]{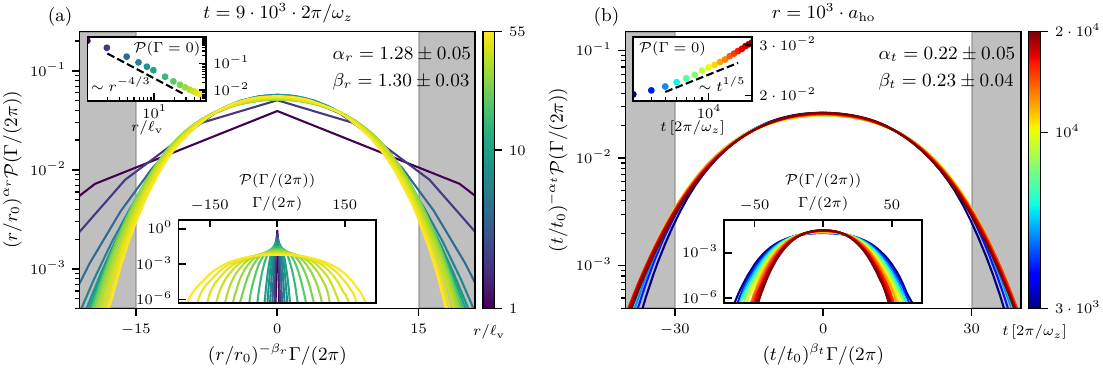}
	\caption{
		(a) Spatially rescaled PDFs at $t=9\cdot10^3(2\pi)/\omega_z$ for $r/\ell_\mathrm{v} \in [1,55]$, where $55\,\ell_{\mathrm{v}}\approx3\cdot10^3\aho$, i.e., the upper boundary of the inertial range, a reference length $r_0=10\,\ell_\mathrm{v}$, and a cutoff $\Gamma_\Lambda/(2\pi)=15$.
		In the upper-left inset the spatial decay of $\mathcal{P}(\Gamma=0)$ is found to closely follow a $\sim r^{-4/3}$ power-law.
		For the spatial scaling exponents we extract $\alpha_r=1.28\pm0.05$ and $\beta_r=1.30\pm0.03$ which both, within errors, agree with $\lambda_1=4/3$ for the IEC.
		For small $r/\ell_\mathrm{v}$ we find worse agreement with the rescaled PDF, which improves once the lower boundary of the inertial range at $r/\ell_\mathrm{v}\approx 6$ is reached.
		(b) Temporally rescaled PDFs at $r=10^3\aho$ within the inertial range, for $t\in[3\cdot10^3,2\cdot10^4](2\pi)/\omega_z$, i.e., the universal time interval, a reference time $t_0=10^4(2\pi)/\omega_z$, and a cutoff $\Gamma_\Lambda/(2\pi)=30$.
		The temporal growth of $\mathcal{P}(\Gamma=0)$ (upper-left inset ) obeys $\sim t^{1/5}$ and the temporal scaling exponents $\alpha_t=0.22\pm0.05$ and $\beta_t=0.23\pm0.04$ are in agreement with the subdiffusive exponent $\beta=1/5$.
		The lower insets in both panels show the non-rescaled PDFs as in \Fig{fig_11}.
	}
	\label{fig:fig_13}
\end{figure*}
%==================================

%==================================
\begin{figure*}[t]
	\centering
	\includegraphics[width=\textwidth]{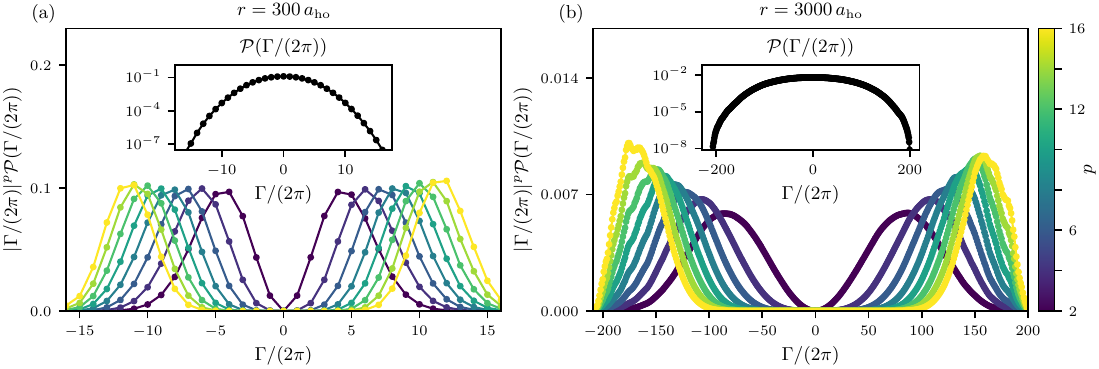}
	\caption{
		Normalized higher-order integrands $|\Gamma/(2\pi)|^p \mathcal{P}(\Gamma/(2\pi))$ of the circulation PDF at (a) $r=300\,\aho$ and (b) $r=3000\,\aho$, i.e., at the boundaries of the inertial range indicated in Fig.~3 of the main text.
		We find all integrands to converge to zero at the maximal $\Gamma$ we can resolve which indicates good statistical convergence up to $p=16$.
		The insets show the PDFs which are approximately Gaussian for small circulations.
	}
	\label{fig:fig_4}
\end{figure*}
%==================================

%==============================================================================
%==============================================================================
\section{Kraichnan-Kolmogorov scaling and intermittency in the higher moments of circulation}
\label{app:Intermittency}
%

%==============================================================================
\subsection{Probability distribution functions}
\label{app:pdf}
%

%==================================
\begin{figure*}[t]
	\centering
	\includegraphics[width=\textwidth]{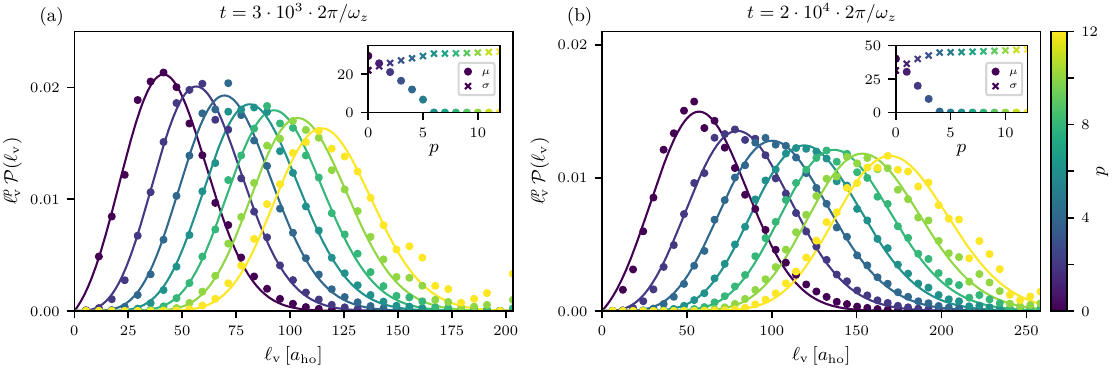}
	\caption{
		The PDF $\mathcal{P}(\ell_\mathrm{v})$ and normalized higher-order integrands $\ell_\mathrm{v}^p \mathcal{P}(\ell_\mathrm{v})$ of the average inter-defect distance $\ell_\mathrm{v}$ at the boundaries of the universal interval $t=3\cdot10^3(2\pi)/\omega_z$ (a) and $t=2\cdot10^4(2\pi)/\omega_z$ (b).
		For the computation we employed $N_\mathrm{v}\approx4.6\cdot10^6$ (a) and $N_\mathrm{v}\approx2.2\cdot10^6$ (b) defects which result in the good statistical convergence up to the $12$th moment.
		The solid lines are Gaussian fits of $A \ell_\mathrm{v}^{1+p} \exp{-(\ell_\mathrm{v}-\mu)^2/\sigma^2}$ which show that the defect distance distribution remains Gaussian up to the highest moment.
		The insets show the fitted values for \(\mu\) and \(\sigma\) which converge around $p\approx5$, indicating that the higher moments are described by the same Gaussian distribution.
	}
	\label{fig:fig_5}
\end{figure*}
%==================================

%==================================
\begin{figure*}[t]
	\centering
	\includegraphics[width=\textwidth]{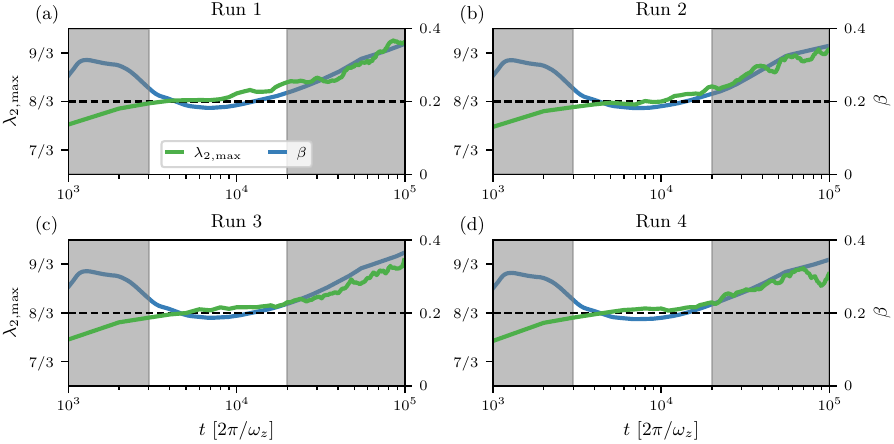}
	\caption{
		Comparison of the maximum local radial logarithmic derivative \(\lambda_{2,\mathrm{max}}\) of the second moment of the velocity circulation \(\Gamma^2(r)\) with the locally fitted temporal scaling exponent \(\beta\).
		Here we show the results for single runs, in contrast to the TWA-averaged result in Fig.~2c of the main text, in order to highlight that the transient behavior is more pronounced since we avoid smoothening through the averaging procedure.
	}
	\label{fig:fig_6}
\end{figure*}
%==================================

For the study of intermittency as shown in Fig.~3 of the main text, we employed higher-order moments of the circulation, Eq.~(1) of the main text, up to order $p=16$ and of the inter-defect distance $\ell_\text{v}$ up to $p=11$.
In this appendix we show the statistical convergence of these moments by computing the probability density functions (PDF) $\mathcal{P}(\Gamma)$ and $\mathcal{P}(\ell_\mathrm{v})$, respectively.
Since the circulation is found to be almost perfectly quantized in integer multiples of $2\pi$, $\Gamma/(2\pi)\in\mathbb{Z}$, we compute PDFs as normalized histograms over integer bins centered at $n/(2\pi)$, $n\in\mathbb{Z}$.
In \Fig{fig_11} we show the obtained PDFs at $t=9\cdot10^3(2\pi)/\omega_z$ for five values of $r/\ell_\text{v}$ chosen (a) $\in\{0.2,0.4,0.6,0.8,1.0\}$ and (b) $\in\{2,4,6,8,10\}$, which exhibit growing probabilities for larger circulations as the radii increase.

Since the PDFs are expected to obey Gaussian statistics for `small' circulations $\Gamma$ we can postulate a scaling hypothesis
\begin{align}
	\label{eq:PDFScalingHypothesis}
	&\mathcal{P}(\Gamma/(2\pi),r,t) = \notag \\
  &\quad(r/r_0)^{-\alpha_r} (t/t_0)^{\alpha_t} \mathcal{P}((r/r_0)^{-\beta_r} (t/t_0)^{\beta_t} \Gamma/(2\pi), r_0, t_0) \,
\end{align}
both in space and time with spatial, $\alpha_r$ and $\beta_r$, and temporal, $\alpha_t$ and $\beta_t$, scaling exponents.
Assuming probability conservation within the interval $[-\Gamma_\Lambda, \Gamma_\Lambda]$ set by the cutoff $\Gamma_\Lambda$, yields the scaling relations $\alpha_t=\beta_t$ and $\alpha_r=\beta_r$.
In \Fig{fig_13} we have tested the scaling hypothesis \eqref{eq:PDFScalingHypothesis} explicitly by rescaling PDFs both spatially and temporally within the inertial range and the universal interval, respectively.
The spatial scaling exponents $\alpha_r=1.28\pm0.05$ and $\beta_r=1.30\pm0.03$ agree well with the expected scaling of $\lambda_1=4/3$ for an inverse cascade.
Likewise, the temporal scaling exponents $\alpha_t=0.22\pm0.05$ and $\beta_t=0.23\pm0.04$ match the subdiffusive exponent $\beta=1/5$, and both scalings satisfy the relation $\alpha_{r,t}\approx\beta_{r,t}$.
This showcases again the simultaneous presence of subdiffusive scaling in the vicinity of the anomalous NTFP and the spatial scaling of an inverse Kraichnan cascade.
The exponents $\alpha_r$ and $\alpha_t$ can also be determined using the probability of zero circulation, as shown in the upper-left insets of \Fig{fig_13}.

In \Fig{fig_4}, normalized integrands for the higher-order moments of the circulation $|\Gamma/(2\pi)|^p \mathcal{P}_\mathrm{PMF}(\Gamma/(2\pi))$ are shown for $p\in\{2,4,\dots,16\}$ at (a) $r=300\,\aho$ and (b) $r=3000\,\aho$, i.e., at the boundaries of the inertial range indicated in Fig.~3 of the main text.
Since the integrands decay to zero for large absolute circulations $|\Gamma|$, we infer statistical convergence of the moments up to the $16$th moment.
The insets show the PDFs which are approximately Gaussian for small circulations, but develop non-Gaussian tails for large $\abs{\Gamma}$.
From the shrinking widths and increasing peak heights of the higher-order integrands in (b), we conclude that circulation becomes increasingly non-Gaussian at larger scales.
This reflects stronger vortex correlations, which give rise to spatial intermittency in collective vortex dynamics, whereas at smaller scales the PDF is consistent with Poissonian distribution of vortices.
This behavior is consistent with the absence of temporal intermittency in the inter-defect distance in Fig.~3b of the main text, which exhibits a Gaussian distribution in \Fig{fig_5}.

%==================================
\begin{figure*}[t]
	\centering
	\includegraphics[width=\textwidth]{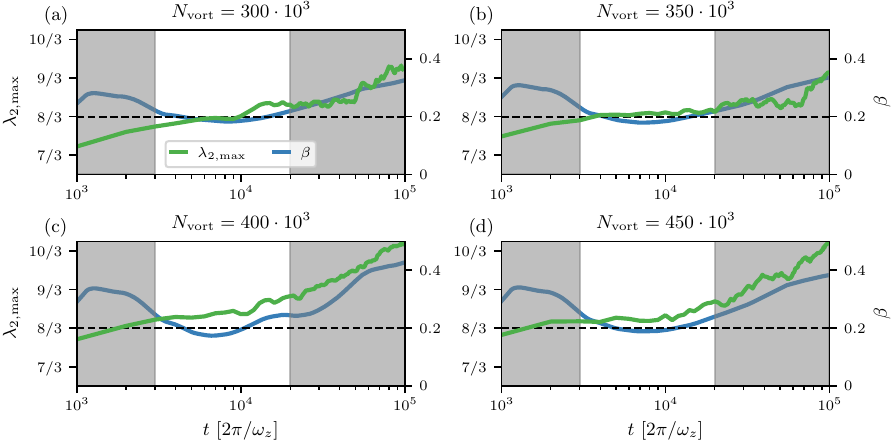}
	\caption{
		To determine the optimal vortex number for observing both coarsening and the IEC in the universal time interval, we show \(\lambda_{2,\mathrm{max}}\) and \(\beta\) for four different initial defect numbers \(N_\mathrm{v} \in \{300,350,400,450\}\cdot10^3\).
		Here we resorted to a reduced grid size of \(8192^2\) for numerical reasons where also the systems length \(L\) and particle number \(N\) have been adopted accordingly.
	}
	\label{fig:fig_7}
\end{figure*}
%==================================

In \Fig{fig_5} we perform a similar analysis for the PDF of the inter-defect distance distribution $\mathcal{P}(\ell_\mathrm{v})$ by computing the normalized integrands $\ell_\mathrm{v}^p\mathcal{P}(\ell_\mathrm{v})$ for even $p$ up to $p=12$.
For large $\ell_\mathrm{v}$ we find the integrands to approximately converge to zero and hence good statistical convergence of the moments computed in Fig.~3 of the main text.
We further fitted Gaussians of the form $A \ell_\mathrm{v}^{1+p} \exp{-(\ell_\mathrm{v}-\mu)^2/\sigma^2}$ to the integrands up to the maximum moment, showcasing that the distribution remains Gaussian.
The insets show the fitted $\mu$ and $\sigma$, which converge around $p\approx5$, and hence, all higher moments can be described by the same Gaussian distribution and no `intermittent' deviations are observed.
The deviation of $\mu$ from zero at small $p$ is attributed to the minimal vortex distance $\sim7a_\mathrm{ho}$ which we employ in order to avoid vortex overcounting in very close dipoles.
Hence we find similar vortex statistics as when dynamically cooling a Bose gas below the second-order condensation phase transition with ensuing vortex creation through the Kibble-Zurek mechanism \cite{Thudiyangal2024}.

%==============================================================================
\subsection{Single-run comparisons of \(\lambda_{2,\mathrm{max}}\) and \(\beta\)}
\label{app:SingleRun}

In this appendix we provide more details on the observed transient behavior of the maximal local slope \(\lambda_{2,\mathrm{max}}\) around \(8/3\) when the system undergoes universal dynamics characterized by coarsening with an exponent \(\beta=1/5\).
Fig.~2c of the main text shows the result we obtained after averaging over twenty individual runs, each starting with \(N_\mathrm{v}=1.4\cdot10^6\) defects.
We complement this with \Fig{fig_6}, showing the same quantities extracted from four individual runs of the simulation. 
These confirm the strong temporal correlation between universal dynamics close to the anomalous NTFP and spatial scaling according to an inverse Kraichnan-Kolmogorov cascade.
All panels show a similar evolution of the temporal exponent \(\beta\) extracted from the decrease of the defect number.
Regarding the maximum local logarithmic radial derivative \(\lambda_{2,\mathrm{max}}\) we find, however, deviations on the single-run level.
In run 1 (a) and run 2 (b), the value \(8/3\) is reached right at the onset of the universal interval at \(t\approx3\cdot10^3(2\pi)/\omega_z\) and remains very close for less than half an order of magnitude in time.
In contrast, run 3 (c) and run 4 (d) show similar early-time behavior but settle to exponents slightly larger than \(8/3\), where they remain stable until the end of the universal interval at \(t\approx2\cdot10^4(2\pi)/\omega_z\).
These small variations between individual runs can be attributed to the randomness of the initial condition but have the disadvantageous effect that the data for \(\lambda_{2,\mathrm{max}}\) is smeared out in the TWA average.
For this reason we show the individual runs to highlight the character of the transient behavior.

%==============================================================================
\subsection{Optimizing the number of defects}
\label{app:DefectNumber}

In the main text we have set the initial defect number to \(N_\mathrm{v}=1.4\cdot10^6\), as we will further motivate in this appendix.
To this end we performed single-run evolutions on a smaller \(8192^2\) numerical grid, where we reduced the system length to \(L/2\) and the particle number to \(N/4\) in order to downsize the computational cost.
\Fig{fig_7} shows the same quantities as in Fig.~2c of the main text and \Fig{fig_6} for the smaller system, with four different initial defect numbers \(N_\mathrm{v} \in \{300,350,400,450\}\cdot10^3\).
For the exponent \(\beta\) we find that at early times \(t\lesssim3\cdot10^3(2\pi)/\omega_z\), increasing the vortex number leads to larger values, which can be explained by the initial overdensity of defects.
During the universal interval, $\beta$ nevertheless approaches the universal value $\approx1/5$, before departing to larger exponents at \(t\gtrsim2\cdot10^4(2\pi)/\omega_z\), which also yields larger exponents for larger vortex numbers.
This is explained by the increased sound created on the background condensate, which then exerts friction on the defects, facilitating the transition to faster coarsening \cite{Rasch:2025kna}.

For the circulation exponent, we find an earlier approach of Kraichnan-Kolmogorov value with increasing defect number which, however, also results in an earlier departure.
The optimum balance between universal dynamics -- best observed for smaller vortex numbers -- and the transient behavior of the Kraichnan IEC -- best observed for larger vortex number -- is around \(N_\mathrm{v}=350\cdot10^3\), cf.~\Fig{fig_7}c.
Multiplying this value by four results in the number used in the main text for the system of fourfold size.

%==============================================================================

\end{appendix}

%==============================================================================

\bibliographystyle{apsrev4-2}

\begin{thebibliography}{121}%
\makeatletter
\providecommand \@ifxundefined [1]{%
 \@ifx{#1\undefined}
}%
\providecommand \@ifnum [1]{%
 \ifnum #1\expandafter \@firstoftwo
 \else \expandafter \@secondoftwo
 \fi
}%
\providecommand \@ifx [1]{%
 \ifx #1\expandafter \@firstoftwo
 \else \expandafter \@secondoftwo
 \fi
}%
\providecommand \natexlab [1]{#1}%
\providecommand \enquote  [1]{``#1''}%
\providecommand \bibnamefont  [1]{#1}%
\providecommand \bibfnamefont [1]{#1}%
\providecommand \citenamefont [1]{#1}%
\providecommand \href@noop [0]{\@secondoftwo}%
\providecommand \href [0]{\begingroup \@sanitize@url \@href}%
\providecommand \@href[1]{\@@startlink{#1}\@@href}%
\providecommand \@@href[1]{\endgroup#1\@@endlink}%
\providecommand \@sanitize@url [0]{\catcode `\\12\catcode `\$12\catcode `\&12\catcode `\#12\catcode `\^12\catcode `\_12\catcode `\%12\relax}%
\providecommand \@@startlink[1]{}%
\providecommand \@@endlink[0]{}%
\providecommand \url  [0]{\begingroup\@sanitize@url \@url }%
\providecommand \@url [1]{\endgroup\@href {#1}{\urlprefix }}%
\providecommand \urlprefix  [0]{URL }%
\providecommand \Eprint [0]{\href }%
\providecommand \doibase [0]{https://doi.org/}%
\providecommand \selectlanguage [0]{\@gobble}%
\providecommand \bibinfo  [0]{\@secondoftwo}%
\providecommand \bibfield  [0]{\@secondoftwo}%
\providecommand \translation [1]{[#1]}%
\providecommand \BibitemOpen [0]{}%
\providecommand \bibitemStop [0]{}%
\providecommand \bibitemNoStop [0]{.\EOS\space}%
\providecommand \EOS [0]{\spacefactor3000\relax}%
\providecommand \BibitemShut  [1]{\csname bibitem#1\endcsname}%
\let\auto@bib@innerbib\@empty
%</preamble>
\bibitem [{\citenamefont {Frisch}(1995)}]{Frisch1995a}%
  \BibitemOpen
  \bibfield  {author} {\bibinfo {author} {\bibfnamefont {U.}~\bibnamefont {Frisch}},\ }\href {https://doi.org/10.1017/CBO9781139170666} {\emph {\bibinfo {title} {Turbulence: The Legacy of A. N. Kolmogorov}}}\ (\bibinfo  {publisher} {CUP, Cambridge},\ \bibinfo {year} {1995})\BibitemShut {NoStop}%
\bibitem [{\citenamefont {{Kolmogorov}}(1991{\natexlab{a}})}]{Kolmogorov1941a.RSPSA.434.9}%
  \BibitemOpen
  \bibfield  {author} {\bibinfo {author} {\bibfnamefont {A.~N.}\ \bibnamefont {{Kolmogorov}}},\ }\href {https://doi.org/10.1098/rspa.1991.0075} {\bibfield  {journal} {\bibinfo  {journal} {Proc. R. Soc. Lond. A}\ }\textbf {\bibinfo {volume} {434}},\ \bibinfo {pages} {9} (\bibinfo {year} {1991}{\natexlab{a}})},\ \bibinfo {note} {[Dokl. Akad. Nauk SSSR {\bf 30}, 301 (1941)]}\BibitemShut {NoStop}%
\bibitem [{\citenamefont {{Kolmogorov}}()}]{Kolmogorov1941b.RussMath20thC.323}%
  \BibitemOpen
  \bibfield  {author} {\bibinfo {author} {\bibfnamefont {A.~N.}\ \bibnamefont {{Kolmogorov}}},\ }\bibinfo {title} {Dissipation of energy in the locally isotropic turbulence},\ in\ \href {https://doi.org/10.1142/9789812779212_0014} {\emph {\bibinfo {booktitle} {Russian Mathematicians in the 20th Century}}},\ \bibinfo {editor} {edited by\ \bibinfo {editor} {\bibfnamefont {Y.}~\bibnamefont {Sinai}}},\ pp.\ \bibinfo {pages} {323--343},\ \bibinfo {note} {[Dokl. Akad. Nauk SSSR {\bf 31}, 538 (1941)]}\BibitemShut {NoStop}%
\bibitem [{\citenamefont {{Kolmogorov}}(1991{\natexlab{b}})}]{Kolmogorov1941c.RSPSA.434.15}%
  \BibitemOpen
  \bibfield  {author} {\bibinfo {author} {\bibfnamefont {A.~N.}\ \bibnamefont {{Kolmogorov}}},\ }\href {https://doi.org/10.1098/rspa.1991.0076} {\bibfield  {journal} {\bibinfo  {journal} {Proc. R. Soc. Lond. A}\ }\textbf {\bibinfo {volume} {434}},\ \bibinfo {pages} {15} (\bibinfo {year} {1991}{\natexlab{b}})},\ \bibinfo {note} {[Dokl. Akad. Nauk SSSR {\bf 32}, 16 (1941)]}\BibitemShut {NoStop}%
\bibitem [{\citenamefont {Obukhov}(1941{\natexlab{a}})}]{Obukhov1941a}%
  \BibitemOpen
  \bibfield  {author} {\bibinfo {author} {\bibfnamefont {A.~M.}\ \bibnamefont {Obukhov}},\ }\href@noop {} {\bibfield  {journal} {\bibinfo  {journal} {Izv. Akad. Nauk SSSR, Ser. Geogr. Geofiz.}\ }\textbf {\bibinfo {volume} {5}},\ \bibinfo {pages} {453} (\bibinfo {year} {1941}{\natexlab{a}})}\BibitemShut {NoStop}%
\bibitem [{\citenamefont {Obukhov}(1941{\natexlab{b}})}]{Obukhov1941b}%
  \BibitemOpen
  \bibfield  {author} {\bibinfo {author} {\bibfnamefont {A.~M.}\ \bibnamefont {Obukhov}},\ }\href@noop {} {\bibfield  {journal} {\bibinfo  {journal} {Dokl. Akad. Nauk SSSR}\ }\textbf {\bibinfo {volume} {32}},\ \bibinfo {pages} {22} (\bibinfo {year} {1941}{\natexlab{b}})}\BibitemShut {NoStop}%
\bibitem [{\citenamefont {{Bodenschatz}}\ and\ \citenamefont {{Eckert}}(2011)}]{Bodenschatz2011a}%
  \BibitemOpen
  \bibfield  {author} {\bibinfo {author} {\bibfnamefont {E.}~\bibnamefont {{Bodenschatz}}}\ and\ \bibinfo {author} {\bibfnamefont {M.}~\bibnamefont {{Eckert}}},\ }in\ \href {https://doi.org/10.48550/arXiv.1107.4729} {\emph {\bibinfo {booktitle} {A Voyage Through Turbulence}}},\ \bibinfo {editor} {edited by\ \bibinfo {editor} {\bibfnamefont {P.}~\bibnamefont {Davidson}}, \bibinfo {editor} {\bibfnamefont {Y.}~\bibnamefont {Kaneda}}, \bibinfo {editor} {\bibfnamefont {K.}~\bibnamefont {Moffatt}},\ and\ \bibinfo {editor} {\bibfnamefont {K.}~\bibnamefont {Sreenivasan}}}\ (\bibinfo  {publisher} {CUP, Cambridge},\ \bibinfo {year} {2011})\BibitemShut {NoStop}%
\bibitem [{\citenamefont {Kraichnan}(1967)}]{Kraichnan1967a}%
  \BibitemOpen
  \bibfield  {author} {\bibinfo {author} {\bibfnamefont {R.}~\bibnamefont {Kraichnan}},\ }\href {https://doi.org/10.1063/1.1762301} {\bibfield  {journal} {\bibinfo  {journal} {Phys. Fluids}\ }\textbf {\bibinfo {volume} {10}},\ \bibinfo {pages} {1417} (\bibinfo {year} {1967})}\BibitemShut {NoStop}%
\bibitem [{\citenamefont {Leith}(1968)}]{Leith1968a.PhysFl11.671}%
  \BibitemOpen
  \bibfield  {author} {\bibinfo {author} {\bibfnamefont {C.~E.}\ \bibnamefont {Leith}},\ }\href {https://doi.org/10.1063/1.1691968} {\bibfield  {journal} {\bibinfo  {journal} {Phys. Fluids}\ }\textbf {\bibinfo {volume} {11}},\ \bibinfo {pages} {671} (\bibinfo {year} {1968})}\BibitemShut {NoStop}%
\bibitem [{\citenamefont {Batchelor}(1969)}]{Batchelor1969a.PhysFl12.II-233}%
  \BibitemOpen
  \bibfield  {author} {\bibinfo {author} {\bibfnamefont {G.~K.}\ \bibnamefont {Batchelor}},\ }\href {https://doi.org/10.1063/1.1692443} {\bibfield  {journal} {\bibinfo  {journal} {Phys. Fluids}\ }\textbf {\bibinfo {volume} {12}},\ \bibinfo {pages} {II 233} (\bibinfo {year} {1969})}\BibitemShut {NoStop}%
\bibitem [{\citenamefont {Volovik}(2004)}]{Volovik2004a}%
  \BibitemOpen
  \bibfield  {author} {\bibinfo {author} {\bibfnamefont {G.}~\bibnamefont {Volovik}},\ }\href {https://doi.org/10.1023/B:JOLT.0000041269.56070.2d} {\bibfield  {journal} {\bibinfo  {journal} {J. Low Temp. Phys.}\ }\textbf {\bibinfo {volume} {136}},\ \bibinfo {pages} {309} (\bibinfo {year} {2004})}\BibitemShut {NoStop}%
\bibitem [{\citenamefont {Tsubota}(2008)}]{Tsubota2008a}%
  \BibitemOpen
  \bibfield  {author} {\bibinfo {author} {\bibfnamefont {M.}~\bibnamefont {Tsubota}},\ }\href {https://doi.org/10.1143/JPSJ.77.111006} {\bibfield  {journal} {\bibinfo  {journal} {J. Phys. Soc. Jpn.}\ }\textbf {\bibinfo {volume} {77}},\ \bibinfo {pages} {111006} (\bibinfo {year} {2008})}\BibitemShut {NoStop}%
\bibitem [{\citenamefont {Barenghi}\ \emph {et~al.}(2014)\citenamefont {Barenghi}, \citenamefont {Skrbek},\ and\ \citenamefont {Sreenivasan}}]{Barenghi2014a.PNAS111.4647}%
  \BibitemOpen
  \bibfield  {author} {\bibinfo {author} {\bibfnamefont {C.~F.}\ \bibnamefont {Barenghi}}, \bibinfo {author} {\bibfnamefont {L.}~\bibnamefont {Skrbek}},\ and\ \bibinfo {author} {\bibfnamefont {K.~R.}\ \bibnamefont {Sreenivasan}},\ }\href {https://doi.org/10.1073/pnas.1400033111} {\bibfield  {journal} {\bibinfo  {journal} {PNAS}\ }\textbf {\bibinfo {volume} {111}},\ \bibinfo {pages} {4647} (\bibinfo {year} {2014})}\BibitemShut {NoStop}%
\bibitem [{\citenamefont {Tsatsos}\ \emph {et~al.}(2016)\citenamefont {Tsatsos}, \citenamefont {Tavares}, \citenamefont {Cidrim}, \citenamefont {Fritsch}, \citenamefont {Caracanhas}, \citenamefont {{dos Santos}}, \citenamefont {Barenghi},\ and\ \citenamefont {Bagnato}}]{Tsatsos2016a}%
  \BibitemOpen
  \bibfield  {author} {\bibinfo {author} {\bibfnamefont {M.~C.}\ \bibnamefont {Tsatsos}}, \bibinfo {author} {\bibfnamefont {P.~E.}\ \bibnamefont {Tavares}}, \bibinfo {author} {\bibfnamefont {A.}~\bibnamefont {Cidrim}}, \bibinfo {author} {\bibfnamefont {A.~R.}\ \bibnamefont {Fritsch}}, \bibinfo {author} {\bibfnamefont {M.~A.}\ \bibnamefont {Caracanhas}}, \bibinfo {author} {\bibfnamefont {F.~E.~A.}\ \bibnamefont {{dos Santos}}}, \bibinfo {author} {\bibfnamefont {C.~F.}\ \bibnamefont {Barenghi}},\ and\ \bibinfo {author} {\bibfnamefont {V.~S.}\ \bibnamefont {Bagnato}},\ }\href {https://doi.org/10.1016/j.physrep.2016.02.003} {\bibfield  {journal} {\bibinfo  {journal} {Phys. Rep.}\ }\textbf {\bibinfo {volume} {622}},\ \bibinfo {pages} {1} (\bibinfo {year} {2016})}\BibitemShut {NoStop}%
\bibitem [{\citenamefont {Johnstone}\ \emph {et~al.}(2019)\citenamefont {Johnstone}, \citenamefont {Groszek}, \citenamefont {Starkey}, \citenamefont {Billington}, \citenamefont {Simula},\ and\ \citenamefont {Helmerson}}]{Johnstone2019a}%
  \BibitemOpen
  \bibfield  {author} {\bibinfo {author} {\bibfnamefont {S.~P.}\ \bibnamefont {Johnstone}}, \bibinfo {author} {\bibfnamefont {A.~J.}\ \bibnamefont {Groszek}}, \bibinfo {author} {\bibfnamefont {P.~T.}\ \bibnamefont {Starkey}}, \bibinfo {author} {\bibfnamefont {C.~J.}\ \bibnamefont {Billington}}, \bibinfo {author} {\bibfnamefont {T.~P.}\ \bibnamefont {Simula}},\ and\ \bibinfo {author} {\bibfnamefont {K.}~\bibnamefont {Helmerson}},\ }\href {https://doi.org/10.1126/science.aat5793} {\bibfield  {journal} {\bibinfo  {journal} {Science}\ }\textbf {\bibinfo {volume} {364}},\ \bibinfo {pages} {1267} (\bibinfo {year} {2019})}\BibitemShut {NoStop}%
\bibitem [{\citenamefont {Gauthier}\ \emph {et~al.}(2019)\citenamefont {Gauthier}, \citenamefont {Reeves}, \citenamefont {Yu}, \citenamefont {Bradley}, \citenamefont {Baker}, \citenamefont {Bell}, \citenamefont {Rubinsztein-Dunlop}, \citenamefont {Davis},\ and\ \citenamefont {Neely}}]{Gauthier2019a}%
  \BibitemOpen
  \bibfield  {author} {\bibinfo {author} {\bibfnamefont {G.}~\bibnamefont {Gauthier}}, \bibinfo {author} {\bibfnamefont {M.~T.}\ \bibnamefont {Reeves}}, \bibinfo {author} {\bibfnamefont {X.}~\bibnamefont {Yu}}, \bibinfo {author} {\bibfnamefont {A.~S.}\ \bibnamefont {Bradley}}, \bibinfo {author} {\bibfnamefont {M.~A.}\ \bibnamefont {Baker}}, \bibinfo {author} {\bibfnamefont {T.~A.}\ \bibnamefont {Bell}}, \bibinfo {author} {\bibfnamefont {H.}~\bibnamefont {Rubinsztein-Dunlop}}, \bibinfo {author} {\bibfnamefont {M.~J.}\ \bibnamefont {Davis}},\ and\ \bibinfo {author} {\bibfnamefont {T.~W.}\ \bibnamefont {Neely}},\ }\href {https://doi.org/10.1126/science.aat5718} {\bibfield  {journal} {\bibinfo  {journal} {Science}\ }\textbf {\bibinfo {volume} {364}},\ \bibinfo {pages} {1264} (\bibinfo {year} {2019})}\BibitemShut {NoStop}%
\bibitem [{\citenamefont {Sachkou}\ \emph {et~al.}(2019)\citenamefont {Sachkou}, \citenamefont {Baker}, \citenamefont {Harris}, \citenamefont {Stockdale}, \citenamefont {Forstner}, \citenamefont {Reeves}, \citenamefont {He}, \citenamefont {McAuslan}, \citenamefont {Bradley}, \citenamefont {Davis},\ and\ \citenamefont {Bowen}}]{Sachkou2019a.Science366.1480}%
  \BibitemOpen
  \bibfield  {author} {\bibinfo {author} {\bibfnamefont {Y.~P.}\ \bibnamefont {Sachkou}}, \bibinfo {author} {\bibfnamefont {C.~G.}\ \bibnamefont {Baker}}, \bibinfo {author} {\bibfnamefont {G.~I.}\ \bibnamefont {Harris}}, \bibinfo {author} {\bibfnamefont {O.~R.}\ \bibnamefont {Stockdale}}, \bibinfo {author} {\bibfnamefont {S.}~\bibnamefont {Forstner}}, \bibinfo {author} {\bibfnamefont {M.~T.}\ \bibnamefont {Reeves}}, \bibinfo {author} {\bibfnamefont {X.}~\bibnamefont {He}}, \bibinfo {author} {\bibfnamefont {D.~L.}\ \bibnamefont {McAuslan}}, \bibinfo {author} {\bibfnamefont {A.~S.}\ \bibnamefont {Bradley}}, \bibinfo {author} {\bibfnamefont {M.~J.}\ \bibnamefont {Davis}},\ and\ \bibinfo {author} {\bibfnamefont {W.~P.}\ \bibnamefont {Bowen}},\ }\href {https://doi.org/10.1126/science.aaw9229} {\bibfield  {journal} {\bibinfo  {journal} {Science}\ }\textbf {\bibinfo {volume} {366}},\ \bibinfo {pages} {1480} (\bibinfo {year} {2019})}\BibitemShut {NoStop}%
\bibitem [{\citenamefont {George}(1992)}]{George1992a.PhysFl4.1492}%
  \BibitemOpen
  \bibfield  {author} {\bibinfo {author} {\bibfnamefont {W.~K.}\ \bibnamefont {George}},\ }\href {https://doi.org/10.1063/1.858423} {\bibfield  {journal} {\bibinfo  {journal} {Phys. Fluids A}\ }\textbf {\bibinfo {volume} {4}},\ \bibinfo {pages} {1492} (\bibinfo {year} {1992})}\BibitemShut {NoStop}%
\bibitem [{\citenamefont {Eyink}\ and\ \citenamefont {Thomson}(2000)}]{Eyink2000a.PhysFl12.477}%
  \BibitemOpen
  \bibfield  {author} {\bibinfo {author} {\bibfnamefont {G.~L.}\ \bibnamefont {Eyink}}\ and\ \bibinfo {author} {\bibfnamefont {D.~J.}\ \bibnamefont {Thomson}},\ }\href {https://doi.org/10.1063/1.870279} {\bibfield  {journal} {\bibinfo  {journal} {Phys. Fluids}\ }\textbf {\bibinfo {volume} {12}},\ \bibinfo {pages} {477} (\bibinfo {year} {2000})}\BibitemShut {NoStop}%
\bibitem [{\citenamefont {Carnevale}\ \emph {et~al.}(1991)\citenamefont {Carnevale}, \citenamefont {McWilliams}, \citenamefont {Pomeau}, \citenamefont {Weiss},\ and\ \citenamefont {Young}}]{Carnevale1991a.PhysRevLett.66.2735}%
  \BibitemOpen
  \bibfield  {author} {\bibinfo {author} {\bibfnamefont {G.~F.}\ \bibnamefont {Carnevale}}, \bibinfo {author} {\bibfnamefont {J.~C.}\ \bibnamefont {McWilliams}}, \bibinfo {author} {\bibfnamefont {Y.}~\bibnamefont {Pomeau}}, \bibinfo {author} {\bibfnamefont {J.~B.}\ \bibnamefont {Weiss}},\ and\ \bibinfo {author} {\bibfnamefont {W.~R.}\ \bibnamefont {Young}},\ }\href {https://doi.org/10.1103/PhysRevLett.66.2735} {\bibfield  {journal} {\bibinfo  {journal} {Phys. Rev. Lett.}\ }\textbf {\bibinfo {volume} {66}},\ \bibinfo {pages} {2735} (\bibinfo {year} {1991})}\BibitemShut {NoStop}%
\bibitem [{\citenamefont {Matthaeus}\ \emph {et~al.}(1991)\citenamefont {Matthaeus}, \citenamefont {Stribling}, \citenamefont {Martinez}, \citenamefont {Oughton},\ and\ \citenamefont {Montgomery}}]{Matthaeus1991a.PhysicaD51.531}%
  \BibitemOpen
  \bibfield  {author} {\bibinfo {author} {\bibfnamefont {W.}~\bibnamefont {Matthaeus}}, \bibinfo {author} {\bibfnamefont {W.}~\bibnamefont {Stribling}}, \bibinfo {author} {\bibfnamefont {D.}~\bibnamefont {Martinez}}, \bibinfo {author} {\bibfnamefont {S.}~\bibnamefont {Oughton}},\ and\ \bibinfo {author} {\bibfnamefont {D.}~\bibnamefont {Montgomery}},\ }\href {https://doi.org/10.1016/0167-2789(91)90259-C} {\bibfield  {journal} {\bibinfo  {journal} {Physica D}\ }\textbf {\bibinfo {volume} {51}},\ \bibinfo {pages} {531} (\bibinfo {year} {1991})}\BibitemShut {NoStop}%
\bibitem [{\citenamefont {Tabeling}\ \emph {et~al.}(1991)\citenamefont {Tabeling}, \citenamefont {Burkhart}, \citenamefont {Cardoso},\ and\ \citenamefont {Willaime}}]{Tabeling1991a.PhysRevLett.67.3772}%
  \BibitemOpen
  \bibfield  {author} {\bibinfo {author} {\bibfnamefont {P.}~\bibnamefont {Tabeling}}, \bibinfo {author} {\bibfnamefont {S.}~\bibnamefont {Burkhart}}, \bibinfo {author} {\bibfnamefont {O.}~\bibnamefont {Cardoso}},\ and\ \bibinfo {author} {\bibfnamefont {H.}~\bibnamefont {Willaime}},\ }\href {https://doi.org/10.1103/PhysRevLett.67.3772} {\bibfield  {journal} {\bibinfo  {journal} {Phys. Rev. Lett.}\ }\textbf {\bibinfo {volume} {67}},\ \bibinfo {pages} {3772} (\bibinfo {year} {1991})}\BibitemShut {NoStop}%
\bibitem [{\citenamefont {Carnevale}\ \emph {et~al.}(1992)\citenamefont {Carnevale}, \citenamefont {McWilliams}, \citenamefont {Pomeau}, \citenamefont {Weiss},\ and\ \citenamefont {Young}}]{Carnevale1992a.PhysFlA.6.1314.McWilliams:endstate}%
  \BibitemOpen
  \bibfield  {author} {\bibinfo {author} {\bibfnamefont {G.~F.}\ \bibnamefont {Carnevale}}, \bibinfo {author} {\bibfnamefont {J.~C.}\ \bibnamefont {McWilliams}}, \bibinfo {author} {\bibfnamefont {Y.}~\bibnamefont {Pomeau}}, \bibinfo {author} {\bibfnamefont {J.~B.}\ \bibnamefont {Weiss}},\ and\ \bibinfo {author} {\bibfnamefont {W.~R.}\ \bibnamefont {Young}},\ }\href {https://doi.org/10.1063/1.858251} {\bibfield  {journal} {\bibinfo  {journal} {Phys. Fluids A}\ }\textbf {\bibinfo {volume} {4}},\ \bibinfo {pages} {1314} (\bibinfo {year} {1992})}\BibitemShut {NoStop}%
\bibitem [{\citenamefont {Cardoso}\ \emph {et~al.}(1994)\citenamefont {Cardoso}, \citenamefont {Marteau},\ and\ \citenamefont {Tabeling}}]{Cardoso1994a.PhysRevE.49.454}%
  \BibitemOpen
  \bibfield  {author} {\bibinfo {author} {\bibfnamefont {O.}~\bibnamefont {Cardoso}}, \bibinfo {author} {\bibfnamefont {D.}~\bibnamefont {Marteau}},\ and\ \bibinfo {author} {\bibfnamefont {P.}~\bibnamefont {Tabeling}},\ }\href {https://doi.org/10.1103/PhysRevE.49.454} {\bibfield  {journal} {\bibinfo  {journal} {Phys. Rev. E}\ }\textbf {\bibinfo {volume} {49}},\ \bibinfo {pages} {454} (\bibinfo {year} {1994})}\BibitemShut {NoStop}%
\bibitem [{\citenamefont {Bracco}\ \emph {et~al.}(2000)\citenamefont {Bracco}, \citenamefont {McWilliams}, \citenamefont {Murante}, \citenamefont {Provenzale},\ and\ \citenamefont {Weiss}}]{Bracco2000a.PhysFl12.2931}%
  \BibitemOpen
  \bibfield  {author} {\bibinfo {author} {\bibfnamefont {A.}~\bibnamefont {Bracco}}, \bibinfo {author} {\bibfnamefont {J.~C.}\ \bibnamefont {McWilliams}}, \bibinfo {author} {\bibfnamefont {G.}~\bibnamefont {Murante}}, \bibinfo {author} {\bibfnamefont {A.}~\bibnamefont {Provenzale}},\ and\ \bibinfo {author} {\bibfnamefont {J.~B.}\ \bibnamefont {Weiss}},\ }\href {https://doi.org/10.1063/1.1290391} {\bibfield  {journal} {\bibinfo  {journal} {Phys. Fluids}\ }\textbf {\bibinfo {volume} {12}},\ \bibinfo {pages} {2931} (\bibinfo {year} {2000})}\BibitemShut {NoStop}%
\bibitem [{\citenamefont {Iwayama}\ \emph {et~al.}(2002)\citenamefont {Iwayama}, \citenamefont {Shepherd},\ and\ \citenamefont {Watanabe}}]{Iwayama2002a}%
  \BibitemOpen
  \bibfield  {author} {\bibinfo {author} {\bibfnamefont {T.}~\bibnamefont {Iwayama}}, \bibinfo {author} {\bibfnamefont {T.~G.}\ \bibnamefont {Shepherd}},\ and\ \bibinfo {author} {\bibfnamefont {T.}~\bibnamefont {Watanabe}},\ }\href {https://doi.org/10.1017/S0022112001007509} {\bibfield  {journal} {\bibinfo  {journal} {J. Fluid Mech.}\ }\textbf {\bibinfo {volume} {456}},\ \bibinfo {pages} {183} (\bibinfo {year} {2002})}\BibitemShut {NoStop}%
\bibitem [{\citenamefont {Danilov}\ \emph {et~al.}(2002)\citenamefont {Danilov}, \citenamefont {Dolzhanskii}, \citenamefont {Dovzhenko},\ and\ \citenamefont {Krymov}}]{Danilov2002a.PhysRevE.65.036316}%
  \BibitemOpen
  \bibfield  {author} {\bibinfo {author} {\bibfnamefont {S.}~\bibnamefont {Danilov}}, \bibinfo {author} {\bibfnamefont {F.~V.}\ \bibnamefont {Dolzhanskii}}, \bibinfo {author} {\bibfnamefont {V.~A.}\ \bibnamefont {Dovzhenko}},\ and\ \bibinfo {author} {\bibfnamefont {V.~A.}\ \bibnamefont {Krymov}},\ }\href {https://doi.org/10.1103/PhysRevE.65.036316} {\bibfield  {journal} {\bibinfo  {journal} {Phys. Rev. E}\ }\textbf {\bibinfo {volume} {65}},\ \bibinfo {pages} {036316} (\bibinfo {year} {2002})}\BibitemShut {NoStop}%
\bibitem [{\citenamefont {van Bokhoven}\ \emph {et~al.}(2007)\citenamefont {van Bokhoven}, \citenamefont {Trieling}, \citenamefont {Clercx},\ and\ \citenamefont {van Heijst}}]{Bokhoven2007a.PhysFl19.046601}%
  \BibitemOpen
  \bibfield  {author} {\bibinfo {author} {\bibfnamefont {L.~J.~A.}\ \bibnamefont {van Bokhoven}}, \bibinfo {author} {\bibfnamefont {R.~R.}\ \bibnamefont {Trieling}}, \bibinfo {author} {\bibfnamefont {H.~J.~H.}\ \bibnamefont {Clercx}},\ and\ \bibinfo {author} {\bibfnamefont {G.~J.~F.}\ \bibnamefont {van Heijst}},\ }\href {https://doi.org/10.1063/1.2716785} {\bibfield  {journal} {\bibinfo  {journal} {Phys. Fluids}\ }\textbf {\bibinfo {volume} {19}},\ \bibinfo {pages} {046601} (\bibinfo {year} {2007})}\BibitemShut {NoStop}%
\bibitem [{\citenamefont {Dritschel}\ \emph {et~al.}(2008)\citenamefont {Dritschel}, \citenamefont {Scott}, \citenamefont {Macaskill}, \citenamefont {Gottwald},\ and\ \citenamefont {Tran}}]{Dritschel2008a.PhysRevLett.101.094501}%
  \BibitemOpen
  \bibfield  {author} {\bibinfo {author} {\bibfnamefont {D.~G.}\ \bibnamefont {Dritschel}}, \bibinfo {author} {\bibfnamefont {R.~K.}\ \bibnamefont {Scott}}, \bibinfo {author} {\bibfnamefont {C.}~\bibnamefont {Macaskill}}, \bibinfo {author} {\bibfnamefont {G.~A.}\ \bibnamefont {Gottwald}},\ and\ \bibinfo {author} {\bibfnamefont {C.~V.}\ \bibnamefont {Tran}},\ }\href {https://doi.org/10.1103/PhysRevLett.101.094501} {\bibfield  {journal} {\bibinfo  {journal} {Phys. Rev. Lett.}\ }\textbf {\bibinfo {volume} {101}},\ \bibinfo {pages} {094501} (\bibinfo {year} {2008})}\BibitemShut {NoStop}%
\bibitem [{\citenamefont {Sire}\ \emph {et~al.}(2011)\citenamefont {Sire}, \citenamefont {Chavanis},\ and\ \citenamefont {Sopik}}]{Sire2010a.PhysRevE.84.056317}%
  \BibitemOpen
  \bibfield  {author} {\bibinfo {author} {\bibfnamefont {C.}~\bibnamefont {Sire}}, \bibinfo {author} {\bibfnamefont {P.-H.}\ \bibnamefont {Chavanis}},\ and\ \bibinfo {author} {\bibfnamefont {J.}~\bibnamefont {Sopik}},\ }\href {https://doi.org/10.1103/PhysRevE.84.056317} {\bibfield  {journal} {\bibinfo  {journal} {Phys. Rev. E}\ }\textbf {\bibinfo {volume} {84}},\ \bibinfo {pages} {056317} (\bibinfo {year} {2011})}\BibitemShut {NoStop}%
\bibitem [{\citenamefont {Mininni}\ and\ \citenamefont {Pouquet}(2013)}]{Mininni2013a.PhysRevE.87.033002}%
  \BibitemOpen
  \bibfield  {author} {\bibinfo {author} {\bibfnamefont {P.~D.}\ \bibnamefont {Mininni}}\ and\ \bibinfo {author} {\bibfnamefont {A.}~\bibnamefont {Pouquet}},\ }\href {https://doi.org/10.1103/PhysRevE.87.033002} {\bibfield  {journal} {\bibinfo  {journal} {Phys. Rev. E}\ }\textbf {\bibinfo {volume} {87}},\ \bibinfo {pages} {033002} (\bibinfo {year} {2013})}\BibitemShut {NoStop}%
\bibitem [{\citenamefont {Fang}\ and\ \citenamefont {Ouellette}(2017)}]{Fang2017a.PhysFl29.111105}%
  \BibitemOpen
  \bibfield  {author} {\bibinfo {author} {\bibfnamefont {L.}~\bibnamefont {Fang}}\ and\ \bibinfo {author} {\bibfnamefont {N.~T.}\ \bibnamefont {Ouellette}},\ }\href {https://doi.org/10.1063/1.4996776} {\bibfield  {journal} {\bibinfo  {journal} {Phys. Fluids}\ }\textbf {\bibinfo {volume} {29}},\ \bibinfo {pages} {111105} (\bibinfo {year} {2017})}\BibitemShut {NoStop}%
\bibitem [{\citenamefont {Chu}\ and\ \citenamefont {Williams}(2001)}]{Chu2001a}%
  \BibitemOpen
  \bibfield  {author} {\bibinfo {author} {\bibfnamefont {H.-C.}\ \bibnamefont {Chu}}\ and\ \bibinfo {author} {\bibfnamefont {G.~A.}\ \bibnamefont {Williams}},\ }\href {https://doi.org/10.1103/PhysRevLett.86.2585} {\bibfield  {journal} {\bibinfo  {journal} {Phys. Rev. Lett.}\ }\textbf {\bibinfo {volume} {86}},\ \bibinfo {pages} {2585} (\bibinfo {year} {2001})}\BibitemShut {NoStop}%
\bibitem [{\citenamefont {Walmsley}\ and\ \citenamefont {Golov}(2008)}]{Walmsley2008aNoArxiv}%
  \BibitemOpen
  \bibfield  {author} {\bibinfo {author} {\bibfnamefont {P.~M.}\ \bibnamefont {Walmsley}}\ and\ \bibinfo {author} {\bibfnamefont {A.~I.}\ \bibnamefont {Golov}},\ }\href {https://doi.org/10.1103/PhysRevLett.100.245301} {\bibfield  {journal} {\bibinfo  {journal} {Phys. Rev. Lett.}\ }\textbf {\bibinfo {volume} {100}},\ \bibinfo {pages} {245301} (\bibinfo {year} {2008})}\BibitemShut {NoStop}%
\bibitem [{\citenamefont {Skrbek}\ and\ \citenamefont {Sreenivasan}(2012)}]{Skrbek2012a.PhysFl24.011301}%
  \BibitemOpen
  \bibfield  {author} {\bibinfo {author} {\bibfnamefont {L.}~\bibnamefont {Skrbek}}\ and\ \bibinfo {author} {\bibfnamefont {K.~R.}\ \bibnamefont {Sreenivasan}},\ }\href {https://doi.org/10.1063/1.3678335} {\bibfield  {journal} {\bibinfo  {journal} {Phys. Fluids}\ }\textbf {\bibinfo {volume} {24}},\ \bibinfo {pages} {011301} (\bibinfo {year} {2012})}\BibitemShut {NoStop}%
\bibitem [{\citenamefont {Baggaley}\ \emph {et~al.}(2012)\citenamefont {Baggaley}, \citenamefont {Barenghi},\ and\ \citenamefont {Sergeev}}]{Baggaley2012b.PhysRevB.85.060501}%
  \BibitemOpen
  \bibfield  {author} {\bibinfo {author} {\bibfnamefont {A.~W.}\ \bibnamefont {Baggaley}}, \bibinfo {author} {\bibfnamefont {C.~F.}\ \bibnamefont {Barenghi}},\ and\ \bibinfo {author} {\bibfnamefont {Y.~A.}\ \bibnamefont {Sergeev}},\ }\href {https://doi.org/10.1103/PhysRevB.85.060501} {\bibfield  {journal} {\bibinfo  {journal} {Phys. Rev. B}\ }\textbf {\bibinfo {volume} {85}},\ \bibinfo {pages} {060501} (\bibinfo {year} {2012})}\BibitemShut {NoStop}%
\bibitem [{\citenamefont {Forrester}\ \emph {et~al.}(2013)\citenamefont {Forrester}, \citenamefont {Chu},\ and\ \citenamefont {Williams}}]{Forrester2013a.PhysRevLett.110.165303}%
  \BibitemOpen
  \bibfield  {author} {\bibinfo {author} {\bibfnamefont {A.}~\bibnamefont {Forrester}}, \bibinfo {author} {\bibfnamefont {H.-C.}\ \bibnamefont {Chu}},\ and\ \bibinfo {author} {\bibfnamefont {G.~A.}\ \bibnamefont {Williams}},\ }\href {https://doi.org/10.1103/PhysRevLett.110.165303} {\bibfield  {journal} {\bibinfo  {journal} {Phys. Rev. Lett.}\ }\textbf {\bibinfo {volume} {110}},\ \bibinfo {pages} {165303} (\bibinfo {year} {2013})}\BibitemShut {NoStop}%
\bibitem [{\citenamefont {Zmeev}\ \emph {et~al.}(2015)\citenamefont {Zmeev}, \citenamefont {Walmsley}, \citenamefont {Golov}, \citenamefont {McClintock}, \citenamefont {Fisher},\ and\ \citenamefont {Vinen}}]{Zmeev2015a}%
  \BibitemOpen
  \bibfield  {author} {\bibinfo {author} {\bibfnamefont {D.~E.}\ \bibnamefont {Zmeev}}, \bibinfo {author} {\bibfnamefont {P.~M.}\ \bibnamefont {Walmsley}}, \bibinfo {author} {\bibfnamefont {A.~I.}\ \bibnamefont {Golov}}, \bibinfo {author} {\bibfnamefont {P.~V.~E.}\ \bibnamefont {McClintock}}, \bibinfo {author} {\bibfnamefont {S.~N.}\ \bibnamefont {Fisher}},\ and\ \bibinfo {author} {\bibfnamefont {W.~F.}\ \bibnamefont {Vinen}},\ }\href {https://doi.org/10.1103/PhysRevLett.115.155303} {\bibfield  {journal} {\bibinfo  {journal} {Phys. Rev. Lett.}\ }\textbf {\bibinfo {volume} {115}},\ \bibinfo {pages} {155303} (\bibinfo {year} {2015})}\BibitemShut {NoStop}%
\bibitem [{\citenamefont {Svistunov}(1991)}]{Svistunov1991a}%
  \BibitemOpen
  \bibfield  {author} {\bibinfo {author} {\bibfnamefont {B.}~\bibnamefont {Svistunov}},\ }\href@noop {} {\bibfield  {journal} {\bibinfo  {journal} {J. Mosc. Phys. Soc.}\ }\textbf {\bibinfo {volume} {1}},\ \bibinfo {pages} {373} (\bibinfo {year} {1991})}\BibitemShut {NoStop}%
\bibitem [{\citenamefont {Kagan}\ \emph {et~al.}(1992)\citenamefont {Kagan}, \citenamefont {Svistunov},\ and\ \citenamefont {Shlyapnikov}}]{Kagan1992a}%
  \BibitemOpen
  \bibfield  {author} {\bibinfo {author} {\bibfnamefont {Y.}~\bibnamefont {Kagan}}, \bibinfo {author} {\bibfnamefont {B.~V.}\ \bibnamefont {Svistunov}},\ and\ \bibinfo {author} {\bibfnamefont {G.~V.}\ \bibnamefont {Shlyapnikov}},\ }\href {http://www.jetp.ac.ru/cgi-bin/e/index/e/74/2/p279?a=list} {\bibfield  {journal} {\bibinfo  {journal} {[Zh. Eksp. Teor. Fiz. 101, 528 (1992)] Sov. Phys. JETP}\ }\textbf {\bibinfo {volume} {74}},\ \bibinfo {pages} {279} (\bibinfo {year} {1992})}\BibitemShut {NoStop}%
\bibitem [{\citenamefont {Kagan}\ and\ \citenamefont {Svistunov}(1994)}]{Kagan1994a}%
  \BibitemOpen
  \bibfield  {author} {\bibinfo {author} {\bibfnamefont {Y.}~\bibnamefont {Kagan}}\ and\ \bibinfo {author} {\bibfnamefont {B.~V.}\ \bibnamefont {Svistunov}},\ }\href {http://www.jetp.ac.ru/cgi-bin/e/index/e/78/2/p187?a=list} {\bibfield  {journal} {\bibinfo  {journal} {[Zh. Eksp. Teor. Fiz. 105, 353 (1994)] Sov. Phys. JETP}\ }\textbf {\bibinfo {volume} {78}},\ \bibinfo {pages} {187} (\bibinfo {year} {1994})}\BibitemShut {NoStop}%
\bibitem [{\citenamefont {Kagan}(1995)}]{Kagan1995a}%
  \BibitemOpen
  \bibfield  {author} {\bibinfo {author} {\bibfnamefont {Y.}~\bibnamefont {Kagan}},\ }\href@noop {} {\emph {\bibinfo {title} {{B}ose-{E}instein Condensation}}}\ (\bibinfo  {publisher} {Cambridge University Press},\ \bibinfo {year} {1995})\ p.\ \bibinfo {pages} {202}\BibitemShut {NoStop}%
\bibitem [{\citenamefont {Semikoz}\ and\ \citenamefont {Tkachev}(1995)}]{Semikoz1995a.PhysRevLett.74.3093}%
  \BibitemOpen
  \bibfield  {author} {\bibinfo {author} {\bibfnamefont {D.~V.}\ \bibnamefont {Semikoz}}\ and\ \bibinfo {author} {\bibfnamefont {I.~I.}\ \bibnamefont {Tkachev}},\ }\href {https://doi.org/10.1103/PhysRevLett.74.3093} {\bibfield  {journal} {\bibinfo  {journal} {Phys. Rev. Lett.}\ }\textbf {\bibinfo {volume} {74}},\ \bibinfo {pages} {3093} (\bibinfo {year} {1995})},\ \Eprint {https://arxiv.org/abs/hep-ph/9409202} {arXiv:hep-ph/9409202 [hep-ph]} \BibitemShut {NoStop}%
\bibitem [{\citenamefont {Semikoz}\ and\ \citenamefont {Tkachev}(1997)}]{Semikoz1997a}%
  \BibitemOpen
  \bibfield  {author} {\bibinfo {author} {\bibfnamefont {D.~V.}\ \bibnamefont {Semikoz}}\ and\ \bibinfo {author} {\bibfnamefont {I.~I.}\ \bibnamefont {Tkachev}},\ }\href {https://doi.org/10.1103/PhysRevD.55.489} {\bibfield  {journal} {\bibinfo  {journal} {Phys. Rev. D}\ }\textbf {\bibinfo {volume} {55}},\ \bibinfo {pages} {489} (\bibinfo {year} {1997})},\ \Eprint {https://arxiv.org/abs/hep-ph/9507306} {hep-ph/9507306} \BibitemShut {NoStop}%
\bibitem [{\citenamefont {Neely}\ \emph {et~al.}(2013)\citenamefont {Neely}, \citenamefont {Bradley}, \citenamefont {Samson}, \citenamefont {Rooney}, \citenamefont {Wright}, \citenamefont {Law}, \citenamefont {Carretero-Gonz\'alez}, \citenamefont {Kevrekidis}, \citenamefont {Davis},\ and\ \citenamefont {Anderson}}]{Neely2013a.PhysRevLett.111.235301}%
  \BibitemOpen
  \bibfield  {author} {\bibinfo {author} {\bibfnamefont {T.~W.}\ \bibnamefont {Neely}}, \bibinfo {author} {\bibfnamefont {A.~S.}\ \bibnamefont {Bradley}}, \bibinfo {author} {\bibfnamefont {E.~C.}\ \bibnamefont {Samson}}, \bibinfo {author} {\bibfnamefont {S.~J.}\ \bibnamefont {Rooney}}, \bibinfo {author} {\bibfnamefont {E.~M.}\ \bibnamefont {Wright}}, \bibinfo {author} {\bibfnamefont {K.~J.~H.}\ \bibnamefont {Law}}, \bibinfo {author} {\bibfnamefont {R.}~\bibnamefont {Carretero-Gonz\'alez}}, \bibinfo {author} {\bibfnamefont {P.~G.}\ \bibnamefont {Kevrekidis}}, \bibinfo {author} {\bibfnamefont {M.~J.}\ \bibnamefont {Davis}},\ and\ \bibinfo {author} {\bibfnamefont {B.~P.}\ \bibnamefont {Anderson}},\ }\href {https://doi.org/10.1103/PhysRevLett.111.235301} {\bibfield  {journal} {\bibinfo  {journal} {Phys. Rev. Lett.}\ }\textbf {\bibinfo {volume} {111}},\ \bibinfo {pages} {235301} (\bibinfo {year} {2013})}\BibitemShut {NoStop}%
\bibitem [{\citenamefont {Billam}\ \emph {et~al.}(2014)\citenamefont {Billam}, \citenamefont {Reeves}, \citenamefont {Anderson},\ and\ \citenamefont {Bradley}}]{Billam2014a.PhysRevLett.112.145301}%
  \BibitemOpen
  \bibfield  {author} {\bibinfo {author} {\bibfnamefont {T.~P.}\ \bibnamefont {Billam}}, \bibinfo {author} {\bibfnamefont {M.~T.}\ \bibnamefont {Reeves}}, \bibinfo {author} {\bibfnamefont {B.~P.}\ \bibnamefont {Anderson}},\ and\ \bibinfo {author} {\bibfnamefont {A.~S.}\ \bibnamefont {Bradley}},\ }\href {https://doi.org/10.1103/PhysRevLett.112.145301} {\bibfield  {journal} {\bibinfo  {journal} {Phys. Rev. Lett.}\ }\textbf {\bibinfo {volume} {112}},\ \bibinfo {pages} {145301} (\bibinfo {year} {2014})}\BibitemShut {NoStop}%
\bibitem [{\citenamefont {Kwon}\ \emph {et~al.}(2014)\citenamefont {Kwon}, \citenamefont {Moon}, \citenamefont {Choi}, \citenamefont {Seo},\ and\ \citenamefont {Shin}}]{Kwon2014a.PhysRevA.90.063627}%
  \BibitemOpen
  \bibfield  {author} {\bibinfo {author} {\bibfnamefont {W.~J.}\ \bibnamefont {Kwon}}, \bibinfo {author} {\bibfnamefont {G.}~\bibnamefont {Moon}}, \bibinfo {author} {\bibfnamefont {J.-y.}\ \bibnamefont {Choi}}, \bibinfo {author} {\bibfnamefont {S.~W.}\ \bibnamefont {Seo}},\ and\ \bibinfo {author} {\bibfnamefont {Y.-i.}\ \bibnamefont {Shin}},\ }\href {https://doi.org/10.1103/PhysRevA.90.063627} {\bibfield  {journal} {\bibinfo  {journal} {Phys. Rev. A}\ }\textbf {\bibinfo {volume} {90}},\ \bibinfo {pages} {063627} (\bibinfo {year} {2014})}\BibitemShut {NoStop}%
\bibitem [{\citenamefont {Stagg}\ \emph {et~al.}(2015)\citenamefont {Stagg}, \citenamefont {Allen}, \citenamefont {Parker},\ and\ \citenamefont {Barenghi}}]{Stagg2015a.PhysRevA.91.013612}%
  \BibitemOpen
  \bibfield  {author} {\bibinfo {author} {\bibfnamefont {G.~W.}\ \bibnamefont {Stagg}}, \bibinfo {author} {\bibfnamefont {A.~J.}\ \bibnamefont {Allen}}, \bibinfo {author} {\bibfnamefont {N.~G.}\ \bibnamefont {Parker}},\ and\ \bibinfo {author} {\bibfnamefont {C.~F.}\ \bibnamefont {Barenghi}},\ }\href {https://doi.org/10.1103/PhysRevA.91.013612} {\bibfield  {journal} {\bibinfo  {journal} {Phys. Rev. A}\ }\textbf {\bibinfo {volume} {91}},\ \bibinfo {pages} {013612} (\bibinfo {year} {2015})}\BibitemShut {NoStop}%
\bibitem [{\citenamefont {Cidrim}\ \emph {et~al.}(2016)\citenamefont {Cidrim}, \citenamefont {dos Santos}, \citenamefont {Galantucci}, \citenamefont {Bagnato},\ and\ \citenamefont {Barenghi}}]{Cidrim2016a.PhysRevA.93.033651}%
  \BibitemOpen
  \bibfield  {author} {\bibinfo {author} {\bibfnamefont {A.}~\bibnamefont {Cidrim}}, \bibinfo {author} {\bibfnamefont {F.~E.~A.}\ \bibnamefont {dos Santos}}, \bibinfo {author} {\bibfnamefont {L.}~\bibnamefont {Galantucci}}, \bibinfo {author} {\bibfnamefont {V.~S.}\ \bibnamefont {Bagnato}},\ and\ \bibinfo {author} {\bibfnamefont {C.~F.}\ \bibnamefont {Barenghi}},\ }\href {https://doi.org/10.1103/PhysRevA.93.033651} {\bibfield  {journal} {\bibinfo  {journal} {Phys. Rev. A}\ }\textbf {\bibinfo {volume} {93}},\ \bibinfo {pages} {033651} (\bibinfo {year} {2016})}\BibitemShut {NoStop}%
\bibitem [{\citenamefont {Groszek}\ \emph {et~al.}(2016)\citenamefont {Groszek}, \citenamefont {Simula}, \citenamefont {Paganin},\ and\ \citenamefont {Helmerson}}]{Groszek2016a.PhysRevA.93.043614}%
  \BibitemOpen
  \bibfield  {author} {\bibinfo {author} {\bibfnamefont {A.~J.}\ \bibnamefont {Groszek}}, \bibinfo {author} {\bibfnamefont {T.~P.}\ \bibnamefont {Simula}}, \bibinfo {author} {\bibfnamefont {D.~M.}\ \bibnamefont {Paganin}},\ and\ \bibinfo {author} {\bibfnamefont {K.}~\bibnamefont {Helmerson}},\ }\href {https://doi.org/10.1103/PhysRevA.93.043614} {\bibfield  {journal} {\bibinfo  {journal} {Phys. Rev. A}\ }\textbf {\bibinfo {volume} {93}},\ \bibinfo {pages} {043614} (\bibinfo {year} {2016})}\BibitemShut {NoStop}%
\bibitem [{\citenamefont {{Seo}}\ \emph {et~al.}(2017)\citenamefont {{Seo}}, \citenamefont {{Ko}}, \citenamefont {{Kim}},\ and\ \citenamefont {{Shin}}}]{Seo2017aNatSciRept7.4587S}%
  \BibitemOpen
  \bibfield  {author} {\bibinfo {author} {\bibfnamefont {S.~W.}\ \bibnamefont {{Seo}}}, \bibinfo {author} {\bibfnamefont {B.}~\bibnamefont {{Ko}}}, \bibinfo {author} {\bibfnamefont {J.~H.}\ \bibnamefont {{Kim}}},\ and\ \bibinfo {author} {\bibfnamefont {Y.}~\bibnamefont {{Shin}}},\ }\href {https://doi.org/10.1038/s41598-017-04122-9} {\bibfield  {journal} {\bibinfo  {journal} {Sci. Rep.}\ }\textbf {\bibinfo {volume} {7}},\ \bibinfo {pages} {4587} (\bibinfo {year} {2017})}\BibitemShut {NoStop}%
\bibitem [{\citenamefont {Baggaley}\ and\ \citenamefont {Barenghi}(2018)}]{Baggaley2018a.PhysRevA.97.033601}%
  \BibitemOpen
  \bibfield  {author} {\bibinfo {author} {\bibfnamefont {A.~W.}\ \bibnamefont {Baggaley}}\ and\ \bibinfo {author} {\bibfnamefont {C.~F.}\ \bibnamefont {Barenghi}},\ }\href {https://doi.org/10.1103/PhysRevA.97.033601} {\bibfield  {journal} {\bibinfo  {journal} {Phys. Rev. A}\ }\textbf {\bibinfo {volume} {97}},\ \bibinfo {pages} {033601} (\bibinfo {year} {2018})}\BibitemShut {NoStop}%
\bibitem [{\citenamefont {M\"uller}\ \emph {et~al.}(2020)\citenamefont {M\"uller}, \citenamefont {Brachet}, \citenamefont {Alexakis},\ and\ \citenamefont {Mininni}}]{Mueller2020a}%
  \BibitemOpen
  \bibfield  {author} {\bibinfo {author} {\bibfnamefont {N.~P.}\ \bibnamefont {M\"uller}}, \bibinfo {author} {\bibfnamefont {M.-E.}\ \bibnamefont {Brachet}}, \bibinfo {author} {\bibfnamefont {A.}~\bibnamefont {Alexakis}},\ and\ \bibinfo {author} {\bibfnamefont {P.~D.}\ \bibnamefont {Mininni}},\ }\href {https://doi.org/10.1103/PhysRevLett.124.134501} {\bibfield  {journal} {\bibinfo  {journal} {Phys. Rev. Lett.}\ }\textbf {\bibinfo {volume} {124}},\ \bibinfo {pages} {134501} (\bibinfo {year} {2020})}\BibitemShut {NoStop}%
\bibitem [{\citenamefont {Kanai}\ and\ \citenamefont {Zhang}(2025)}]{Kanai2025a}%
  \BibitemOpen
  \bibfield  {author} {\bibinfo {author} {\bibfnamefont {T.}~\bibnamefont {Kanai}}\ and\ \bibinfo {author} {\bibfnamefont {C.}~\bibnamefont {Zhang}},\ }\href {https://doi.org/10.1103/6zfl-dc7r} {\bibfield  {journal} {\bibinfo  {journal} {Phys. Rev. A}\ }\textbf {\bibinfo {volume} {112}},\ \bibinfo {pages} {033305} (\bibinfo {year} {2025})}\BibitemShut {NoStop}%
\bibitem [{\citenamefont {Novotný}\ \emph {et~al.}(2025)\citenamefont {Novotný}, \citenamefont {Talíř},\ and\ \citenamefont {Varga}}]{Novotny2025a}%
  \BibitemOpen
  \bibfield  {author} {\bibinfo {author} {\bibfnamefont {F.}~\bibnamefont {Novotný}}, \bibinfo {author} {\bibfnamefont {M.}~\bibnamefont {Talíř}},\ and\ \bibinfo {author} {\bibfnamefont {E.}~\bibnamefont {Varga}},\ }\href {https://arxiv.org/abs/2509.05966} {\bibinfo {title} {Decay of two-dimensional superfluid turbulence over pinning surface}} (\bibinfo {year} {2025}),\ \Eprint {https://arxiv.org/abs/2509.05966} {2509.05966} \BibitemShut {NoStop}%
\bibitem [{\citenamefont {Forrester}\ and\ \citenamefont {Williams}(2014)}]{Forrester2014a.JPhysConfSer568.012031}%
  \BibitemOpen
  \bibfield  {author} {\bibinfo {author} {\bibfnamefont {A.}~\bibnamefont {Forrester}}\ and\ \bibinfo {author} {\bibfnamefont {G.~A.}\ \bibnamefont {Williams}},\ }\href {https://doi.org/10.1088/1742-6596/568/012031} {\bibfield  {journal} {\bibinfo  {journal} {Journal of Physics: Conference Series}\ }\textbf {\bibinfo {volume} {568}},\ \bibinfo {pages} {012031} (\bibinfo {year} {2014})}\BibitemShut {NoStop}%
\bibitem [{\citenamefont {Reeves}\ \emph {et~al.}(2017)\citenamefont {Reeves}, \citenamefont {Billam}, \citenamefont {Yu},\ and\ \citenamefont {Bradley}}]{Reeves2017a.PhysRevLett.119.184502}%
  \BibitemOpen
  \bibfield  {author} {\bibinfo {author} {\bibfnamefont {M.~T.}\ \bibnamefont {Reeves}}, \bibinfo {author} {\bibfnamefont {T.~P.}\ \bibnamefont {Billam}}, \bibinfo {author} {\bibfnamefont {X.}~\bibnamefont {Yu}},\ and\ \bibinfo {author} {\bibfnamefont {A.~S.}\ \bibnamefont {Bradley}},\ }\href {https://doi.org/10.1103/PhysRevLett.119.184502} {\bibfield  {journal} {\bibinfo  {journal} {Phys. Rev. Lett.}\ }\textbf {\bibinfo {volume} {119}},\ \bibinfo {pages} {184502} (\bibinfo {year} {2017})}\BibitemShut {NoStop}%
\bibitem [{\citenamefont {Forrester}\ \emph {et~al.}(2020)\citenamefont {Forrester}, \citenamefont {Chu},\ and\ \citenamefont {Williams}}]{Forrester2020a}%
  \BibitemOpen
  \bibfield  {author} {\bibinfo {author} {\bibfnamefont {A.}~\bibnamefont {Forrester}}, \bibinfo {author} {\bibfnamefont {H.-C.}\ \bibnamefont {Chu}},\ and\ \bibinfo {author} {\bibfnamefont {G.~A.}\ \bibnamefont {Williams}},\ }\href {https://doi.org/10.1103/PhysRevFluids.5.072701} {\bibfield  {journal} {\bibinfo  {journal} {Phys. Rev. Fluids}\ }\textbf {\bibinfo {volume} {5}},\ \bibinfo {pages} {072701} (\bibinfo {year} {2020})}\BibitemShut {NoStop}%
\bibitem [{\citenamefont {Williams}(2022)}]{Williams2022a}%
  \BibitemOpen
  \bibfield  {author} {\bibinfo {author} {\bibfnamefont {G.~A.}\ \bibnamefont {Williams}},\ }\href {https://doi.org/10.1007/s10909-021-02663-y} {\bibfield  {journal} {\bibinfo  {journal} {Journal of Low Temperature Physics}\ }\textbf {\bibinfo {volume} {208}},\ \bibinfo {pages} {394} (\bibinfo {year} {2022})}\BibitemShut {NoStop}%
\bibitem [{\citenamefont {Groszek}\ \emph {et~al.}(2020)\citenamefont {Groszek}, \citenamefont {Davis},\ and\ \citenamefont {Simula}}]{Groszek2020a.SciPostPhys.8.3.039}%
  \BibitemOpen
  \bibfield  {author} {\bibinfo {author} {\bibfnamefont {A.~J.}\ \bibnamefont {Groszek}}, \bibinfo {author} {\bibfnamefont {M.~J.}\ \bibnamefont {Davis}},\ and\ \bibinfo {author} {\bibfnamefont {T.~P.}\ \bibnamefont {Simula}},\ }\href {https://doi.org/10.21468/SciPostPhys.8.3.039} {\bibfield  {journal} {\bibinfo  {journal} {SciPost Phys.}\ }\textbf {\bibinfo {volume} {8}},\ \bibinfo {pages} {039} (\bibinfo {year} {2020})}\BibitemShut {NoStop}%
\bibitem [{\citenamefont {Groszek}\ \emph {et~al.}(2021)\citenamefont {Groszek}, \citenamefont {Comaron}, \citenamefont {Proukakis},\ and\ \citenamefont {Billam}}]{Groszek2021a.PhysRevResearch.3.013212}%
  \BibitemOpen
  \bibfield  {author} {\bibinfo {author} {\bibfnamefont {A.~J.}\ \bibnamefont {Groszek}}, \bibinfo {author} {\bibfnamefont {P.}~\bibnamefont {Comaron}}, \bibinfo {author} {\bibfnamefont {N.~P.}\ \bibnamefont {Proukakis}},\ and\ \bibinfo {author} {\bibfnamefont {T.~P.}\ \bibnamefont {Billam}},\ }\href {https://doi.org/10.1103/PhysRevResearch.3.013212} {\bibfield  {journal} {\bibinfo  {journal} {Phys. Rev. Res.}\ }\textbf {\bibinfo {volume} {3}},\ \bibinfo {pages} {013212} (\bibinfo {year} {2021})}\BibitemShut {NoStop}%
\bibitem [{\citenamefont {Berges}\ \emph {et~al.}(2008)\citenamefont {Berges}, \citenamefont {Rothkopf},\ and\ \citenamefont {Schmidt}}]{Berges2008a}%
  \BibitemOpen
  \bibfield  {author} {\bibinfo {author} {\bibfnamefont {J.}~\bibnamefont {Berges}}, \bibinfo {author} {\bibfnamefont {A.}~\bibnamefont {Rothkopf}},\ and\ \bibinfo {author} {\bibfnamefont {J.}~\bibnamefont {Schmidt}},\ }\href {https://doi.org/10.1103/PhysRevLett.101.041603} {\bibfield  {journal} {\bibinfo  {journal} {Phys. Rev. Lett.}\ }\textbf {\bibinfo {volume} {101}},\ \bibinfo {pages} {041603} (\bibinfo {year} {2008})}\BibitemShut {NoStop}%
\bibitem [{\citenamefont {Berges}\ and\ \citenamefont {Hoffmeister}(2009)}]{Berges2008b}%
  \BibitemOpen
  \bibfield  {author} {\bibinfo {author} {\bibfnamefont {J.}~\bibnamefont {Berges}}\ and\ \bibinfo {author} {\bibfnamefont {G.}~\bibnamefont {Hoffmeister}},\ }\href {https://doi.org/10.1016/j.nuclphysb.2008.12.017} {\bibfield  {journal} {\bibinfo  {journal} {Nucl. Phys. B}\ }\textbf {\bibinfo {volume} {813}},\ \bibinfo {pages} {383} (\bibinfo {year} {2009})}\BibitemShut {NoStop}%
\bibitem [{\citenamefont {Scheppach}\ \emph {et~al.}(2010)\citenamefont {Scheppach}, \citenamefont {Berges},\ and\ \citenamefont {Gasenzer}}]{Scheppach2009a}%
  \BibitemOpen
  \bibfield  {author} {\bibinfo {author} {\bibfnamefont {C.}~\bibnamefont {Scheppach}}, \bibinfo {author} {\bibfnamefont {J.}~\bibnamefont {Berges}},\ and\ \bibinfo {author} {\bibfnamefont {T.}~\bibnamefont {Gasenzer}},\ }\href {https://doi.org/10.1103/PhysRevA.81.033611} {\bibfield  {journal} {\bibinfo  {journal} {Phys. Rev. A}\ }\textbf {\bibinfo {volume} {81}},\ \bibinfo {pages} {033611} (\bibinfo {year} {2010})}\BibitemShut {NoStop}%
\bibitem [{\citenamefont {Pi{\~n}eiro~Orioli}\ \emph {et~al.}(2015)\citenamefont {Pi{\~n}eiro~Orioli}, \citenamefont {Boguslavski},\ and\ \citenamefont {Berges}}]{PineiroOrioli2015a}%
  \BibitemOpen
  \bibfield  {author} {\bibinfo {author} {\bibfnamefont {A.}~\bibnamefont {Pi{\~n}eiro~Orioli}}, \bibinfo {author} {\bibfnamefont {K.}~\bibnamefont {Boguslavski}},\ and\ \bibinfo {author} {\bibfnamefont {J.}~\bibnamefont {Berges}},\ }\href {https://doi.org/10.1103/PhysRevD.92.025041} {\bibfield  {journal} {\bibinfo  {journal} {Phys. Rev. D}\ }\textbf {\bibinfo {volume} {92}},\ \bibinfo {pages} {025041} (\bibinfo {year} {2015})}\BibitemShut {NoStop}%
\bibitem [{\citenamefont {Chantesana}\ \emph {et~al.}(2019)\citenamefont {Chantesana}, \citenamefont {Pi{\~n}eiro~Orioli},\ and\ \citenamefont {Gasenzer}}]{Chantesana2018a}%
  \BibitemOpen
  \bibfield  {author} {\bibinfo {author} {\bibfnamefont {I.}~\bibnamefont {Chantesana}}, \bibinfo {author} {\bibfnamefont {A.}~\bibnamefont {Pi{\~n}eiro~Orioli}},\ and\ \bibinfo {author} {\bibfnamefont {T.}~\bibnamefont {Gasenzer}},\ }\href {https://doi.org/10.1103/PhysRevA.99.043620} {\bibfield  {journal} {\bibinfo  {journal} {Phys. Rev. A}\ }\textbf {\bibinfo {volume} {99}},\ \bibinfo {pages} {043620} (\bibinfo {year} {2019})}\BibitemShut {NoStop}%
\bibitem [{\citenamefont {Gasenzer}\ \emph {et~al.}(2012)\citenamefont {Gasenzer}, \citenamefont {Nowak},\ and\ \citenamefont {Sexty}}]{Gasenzer2011a}%
  \BibitemOpen
  \bibfield  {author} {\bibinfo {author} {\bibfnamefont {T.}~\bibnamefont {Gasenzer}}, \bibinfo {author} {\bibfnamefont {B.}~\bibnamefont {Nowak}},\ and\ \bibinfo {author} {\bibfnamefont {D.}~\bibnamefont {Sexty}},\ }\href {https://doi.org/10.1016/j.physletb.2012.03.031} {\bibfield  {journal} {\bibinfo  {journal} {Phys. Lett. B}\ }\textbf {\bibinfo {volume} {710}},\ \bibinfo {pages} {500} (\bibinfo {year} {2012})}\BibitemShut {NoStop}%
\bibitem [{\citenamefont {Nowak}\ \emph {et~al.}(2012{\natexlab{a}})\citenamefont {Nowak}, \citenamefont {Schole}, \citenamefont {Sexty},\ and\ \citenamefont {Gasenzer}}]{Nowak2011a}%
  \BibitemOpen
  \bibfield  {author} {\bibinfo {author} {\bibfnamefont {B.}~\bibnamefont {Nowak}}, \bibinfo {author} {\bibfnamefont {J.}~\bibnamefont {Schole}}, \bibinfo {author} {\bibfnamefont {D.}~\bibnamefont {Sexty}},\ and\ \bibinfo {author} {\bibfnamefont {T.}~\bibnamefont {Gasenzer}},\ }\href {https://doi.org/10.1103/PhysRevA.85.043627} {\bibfield  {journal} {\bibinfo  {journal} {Phys. Rev. A}\ }\textbf {\bibinfo {volume} {85}},\ \bibinfo {pages} {043627} (\bibinfo {year} {2012}{\natexlab{a}})}\BibitemShut {NoStop}%
\bibitem [{\citenamefont {{Nowak}}\ \emph {et~al.}(2014{\natexlab{a}})\citenamefont {{Nowak}}, \citenamefont {{Schole}},\ and\ \citenamefont {{Gasenzer}}}]{Nowak2012b}%
  \BibitemOpen
  \bibfield  {author} {\bibinfo {author} {\bibfnamefont {B.}~\bibnamefont {{Nowak}}}, \bibinfo {author} {\bibfnamefont {J.}~\bibnamefont {{Schole}}},\ and\ \bibinfo {author} {\bibfnamefont {T.}~\bibnamefont {{Gasenzer}}},\ }\href {https://doi.org/10.1088/1367-2630/16/9/093052} {\bibfield  {journal} {\bibinfo  {journal} {New J. Phys.}\ }\textbf {\bibinfo {volume} {16}},\ \bibinfo {pages} {093052} (\bibinfo {year} {2014}{\natexlab{a}})}\BibitemShut {NoStop}%
\bibitem [{\citenamefont {Schole}\ \emph {et~al.}(2012{\natexlab{a}})\citenamefont {Schole}, \citenamefont {Nowak},\ and\ \citenamefont {Gasenzer}}]{Schole2012a}%
  \BibitemOpen
  \bibfield  {author} {\bibinfo {author} {\bibfnamefont {J.}~\bibnamefont {Schole}}, \bibinfo {author} {\bibfnamefont {B.}~\bibnamefont {Nowak}},\ and\ \bibinfo {author} {\bibfnamefont {T.}~\bibnamefont {Gasenzer}},\ }\href {https://doi.org/10.1103/PhysRevA.86.013624} {\bibfield  {journal} {\bibinfo  {journal} {Phys. Rev. A}\ }\textbf {\bibinfo {volume} {86}},\ \bibinfo {pages} {013624} (\bibinfo {year} {2012}{\natexlab{a}})}\BibitemShut {NoStop}%
\bibitem [{\citenamefont {Gasenzer}\ \emph {et~al.}(2014{\natexlab{a}})\citenamefont {Gasenzer}, \citenamefont {McLerran}, \citenamefont {Pawlowski},\ and\ \citenamefont {Sexty}}]{Gasenzer2013a}%
  \BibitemOpen
  \bibfield  {author} {\bibinfo {author} {\bibfnamefont {T.}~\bibnamefont {Gasenzer}}, \bibinfo {author} {\bibfnamefont {L.}~\bibnamefont {McLerran}}, \bibinfo {author} {\bibfnamefont {J.~M.}\ \bibnamefont {Pawlowski}},\ and\ \bibinfo {author} {\bibfnamefont {D.}~\bibnamefont {Sexty}},\ }\href {https://doi.org/10.1016/j.nuclphysa.2014.07.030} {\bibfield  {journal} {\bibinfo  {journal} {Nucl. Phys. A}\ }\textbf {\bibinfo {volume} {930}},\ \bibinfo {pages} {163} (\bibinfo {year} {2014}{\natexlab{a}})}\BibitemShut {NoStop}%
\bibitem [{\citenamefont {Ewerz}\ \emph {et~al.}(2015{\natexlab{a}})\citenamefont {Ewerz}, \citenamefont {Gasenzer}, \citenamefont {Karl},\ and\ \citenamefont {Samberg}}]{Ewerz2014a}%
  \BibitemOpen
  \bibfield  {author} {\bibinfo {author} {\bibfnamefont {C.}~\bibnamefont {Ewerz}}, \bibinfo {author} {\bibfnamefont {T.}~\bibnamefont {Gasenzer}}, \bibinfo {author} {\bibfnamefont {M.}~\bibnamefont {Karl}},\ and\ \bibinfo {author} {\bibfnamefont {A.}~\bibnamefont {Samberg}},\ }\href {https://doi.org/10.1007/JHEP05(2015)070} {\bibfield  {journal} {\bibinfo  {journal} {J. High Energ. Phys.}\ }\textbf {\bibinfo {volume} {2015}},\ \bibinfo {pages} {70}}\BibitemShut {NoStop}%
\bibitem [{\citenamefont {Karl}\ and\ \citenamefont {Gasenzer}(2017)}]{Karl2017b}%
  \BibitemOpen
  \bibfield  {author} {\bibinfo {author} {\bibfnamefont {M.}~\bibnamefont {Karl}}\ and\ \bibinfo {author} {\bibfnamefont {T.}~\bibnamefont {Gasenzer}},\ }\href {https://doi.org/10.1088/1367-2630/aa7eeb} {\bibfield  {journal} {\bibinfo  {journal} {New J. Phys.}\ }\textbf {\bibinfo {volume} {19}},\ \bibinfo {pages} {093014} (\bibinfo {year} {2017})}\BibitemShut {NoStop}%
\bibitem [{\citenamefont {Deng}\ \emph {et~al.}(2018)\citenamefont {Deng}, \citenamefont {Schlichting}, \citenamefont {Venugopalan},\ and\ \citenamefont {Wang}}]{Deng2018a}%
  \BibitemOpen
  \bibfield  {author} {\bibinfo {author} {\bibfnamefont {J.}~\bibnamefont {Deng}}, \bibinfo {author} {\bibfnamefont {S.}~\bibnamefont {Schlichting}}, \bibinfo {author} {\bibfnamefont {R.}~\bibnamefont {Venugopalan}},\ and\ \bibinfo {author} {\bibfnamefont {Q.}~\bibnamefont {Wang}},\ }\href {https://doi.org/10.1103/PhysRevA.97.053606} {\bibfield  {journal} {\bibinfo  {journal} {Phys. Rev. A}\ }\textbf {\bibinfo {volume} {97}},\ \bibinfo {pages} {053606} (\bibinfo {year} {2018})}\BibitemShut {NoStop}%
\bibitem [{\citenamefont {Schmied}\ \emph {et~al.}(2019{\natexlab{a}})\citenamefont {Schmied}, \citenamefont {Pr\"ufer}, \citenamefont {Oberthaler},\ and\ \citenamefont {Gasenzer}}]{Schmied:2018osf.PhysRevA.99.033611}%
  \BibitemOpen
  \bibfield  {author} {\bibinfo {author} {\bibfnamefont {C.-M.}\ \bibnamefont {Schmied}}, \bibinfo {author} {\bibfnamefont {M.}~\bibnamefont {Pr\"ufer}}, \bibinfo {author} {\bibfnamefont {M.~K.}\ \bibnamefont {Oberthaler}},\ and\ \bibinfo {author} {\bibfnamefont {T.}~\bibnamefont {Gasenzer}},\ }\href {https://doi.org/10.1103/PhysRevA.99.033611} {\bibfield  {journal} {\bibinfo  {journal} {Phys. Rev. A}\ }\textbf {\bibinfo {volume} {99}},\ \bibinfo {pages} {033611} (\bibinfo {year} {2019}{\natexlab{a}})}\BibitemShut {NoStop}%
\bibitem [{\citenamefont {Schmied}\ \emph {et~al.}(2019{\natexlab{b}})\citenamefont {Schmied}, \citenamefont {Gasenzer},\ and\ \citenamefont {Blakie}}]{Schmied2019a}%
  \BibitemOpen
  \bibfield  {author} {\bibinfo {author} {\bibfnamefont {C.~M.}\ \bibnamefont {Schmied}}, \bibinfo {author} {\bibfnamefont {T.}~\bibnamefont {Gasenzer}},\ and\ \bibinfo {author} {\bibfnamefont {P.~B.}\ \bibnamefont {Blakie}},\ }\href {https://doi.org/10.1103/PhysRevA.100.033603} {\bibfield  {journal} {\bibinfo  {journal} {Phys. Rev. A}\ }\textbf {\bibinfo {volume} {100}},\ \bibinfo {pages} {033603} (\bibinfo {year} {2019}{\natexlab{b}})}\BibitemShut {NoStop}%
\bibitem [{\citenamefont {Spitz}\ \emph {et~al.}(2021)\citenamefont {Spitz}, \citenamefont {Berges}, \citenamefont {Oberthaler},\ and\ \citenamefont {Wienhard}}]{Spitz2021a.SciPostPhys11.3.060}%
  \BibitemOpen
  \bibfield  {author} {\bibinfo {author} {\bibfnamefont {D.}~\bibnamefont {Spitz}}, \bibinfo {author} {\bibfnamefont {J.}~\bibnamefont {Berges}}, \bibinfo {author} {\bibfnamefont {M.}~\bibnamefont {Oberthaler}},\ and\ \bibinfo {author} {\bibfnamefont {A.}~\bibnamefont {Wienhard}},\ }\href {https://doi.org/10.21468/scipostphys.11.3.060} {\bibfield  {journal} {\bibinfo  {journal} {SciPost Physics}\ }\textbf {\bibinfo {volume} {11}},\ \bibinfo {pages} {060} (\bibinfo {year} {2021})}\BibitemShut {NoStop}%
\bibitem [{\citenamefont {Siovitz}\ \emph {et~al.}(2023)\citenamefont {Siovitz}, \citenamefont {Lannig}, \citenamefont {Deller}, \citenamefont {Strobel}, \citenamefont {Oberthaler},\ and\ \citenamefont {Gasenzer}}]{Siovitz:2023ius.PhysRevLett.131.183402}%
  \BibitemOpen
  \bibfield  {author} {\bibinfo {author} {\bibfnamefont {I.}~\bibnamefont {Siovitz}}, \bibinfo {author} {\bibfnamefont {S.}~\bibnamefont {Lannig}}, \bibinfo {author} {\bibfnamefont {Y.}~\bibnamefont {Deller}}, \bibinfo {author} {\bibfnamefont {H.}~\bibnamefont {Strobel}}, \bibinfo {author} {\bibfnamefont {M.~K.}\ \bibnamefont {Oberthaler}},\ and\ \bibinfo {author} {\bibfnamefont {T.}~\bibnamefont {Gasenzer}},\ }\href {https://doi.org/10.1103/PhysRevLett.131.183402} {\bibfield  {journal} {\bibinfo  {journal} {Phys. Rev. Lett.}\ }\textbf {\bibinfo {volume} {131}},\ \bibinfo {pages} {183402} (\bibinfo {year} {2023})}\BibitemShut {NoStop}%
\bibitem [{\citenamefont {Huh}\ \emph {et~al.}(2024)\citenamefont {Huh}, \citenamefont {Mukherjee}, \citenamefont {Kwon}, \citenamefont {Seo}, \citenamefont {Hur}, \citenamefont {Mistakidis}, \citenamefont {Sadeghpour},\ and\ \citenamefont {Choi}}]{Huh:2023xso}%
  \BibitemOpen
  \bibfield  {author} {\bibinfo {author} {\bibfnamefont {S.}~\bibnamefont {Huh}}, \bibinfo {author} {\bibfnamefont {K.}~\bibnamefont {Mukherjee}}, \bibinfo {author} {\bibfnamefont {K.}~\bibnamefont {Kwon}}, \bibinfo {author} {\bibfnamefont {J.}~\bibnamefont {Seo}}, \bibinfo {author} {\bibfnamefont {J.}~\bibnamefont {Hur}}, \bibinfo {author} {\bibfnamefont {S.~I.}\ \bibnamefont {Mistakidis}}, \bibinfo {author} {\bibfnamefont {H.~R.}\ \bibnamefont {Sadeghpour}},\ and\ \bibinfo {author} {\bibfnamefont {J.}~\bibnamefont {Choi}},\ }\href {https://doi.org/10.1038/s41567-023-02339-2} {\bibfield  {journal} {\bibinfo  {journal} {Nat. Phys.}\ }\textbf {\bibinfo {volume} {20}},\ \bibinfo {pages} {402} (\bibinfo {year} {2024})}\BibitemShut {NoStop}%
\bibitem [{\citenamefont {Siovitz}\ \emph {et~al.}(2025)\citenamefont {Siovitz}, \citenamefont {Gl\"uck}, \citenamefont {Deller}, \citenamefont {Schmutz}, \citenamefont {Klein}, \citenamefont {Strobel}, \citenamefont {Oberthaler},\ and\ \citenamefont {Gasenzer}}]{Siovitz2025a.PRA112.023304}%
  \BibitemOpen
  \bibfield  {author} {\bibinfo {author} {\bibfnamefont {I.}~\bibnamefont {Siovitz}}, \bibinfo {author} {\bibfnamefont {A.-M.~E.}\ \bibnamefont {Gl\"uck}}, \bibinfo {author} {\bibfnamefont {Y.}~\bibnamefont {Deller}}, \bibinfo {author} {\bibfnamefont {A.}~\bibnamefont {Schmutz}}, \bibinfo {author} {\bibfnamefont {F.}~\bibnamefont {Klein}}, \bibinfo {author} {\bibfnamefont {H.}~\bibnamefont {Strobel}}, \bibinfo {author} {\bibfnamefont {M.~K.}\ \bibnamefont {Oberthaler}},\ and\ \bibinfo {author} {\bibfnamefont {T.}~\bibnamefont {Gasenzer}},\ }\href {https://doi.org/10.1103/df5w-3yfd} {\bibfield  {journal} {\bibinfo  {journal} {Phys. Rev. A}\ }\textbf {\bibinfo {volume} {112}},\ \bibinfo {pages} {023304} (\bibinfo {year} {2025})}\BibitemShut {NoStop}%
\bibitem [{\citenamefont {Noel}\ and\ \citenamefont {Spitz}(2024)}]{Noel2024:PhysRevD.109.056011}%
  \BibitemOpen
  \bibfield  {author} {\bibinfo {author} {\bibfnamefont {V.}~\bibnamefont {Noel}}\ and\ \bibinfo {author} {\bibfnamefont {D.}~\bibnamefont {Spitz}},\ }\href {https://doi.org/10.1103/PhysRevD.109.056011} {\bibfield  {journal} {\bibinfo  {journal} {Phys. Rev. D}\ }\textbf {\bibinfo {volume} {109}},\ \bibinfo {pages} {056011} (\bibinfo {year} {2024})}\BibitemShut {NoStop}%
\bibitem [{\citenamefont {Noel}\ \emph {et~al.}(2025)\citenamefont {Noel}, \citenamefont {Gasenzer},\ and\ \citenamefont {Boguslavski}}]{Noel2025a.PRR7.033220}%
  \BibitemOpen
  \bibfield  {author} {\bibinfo {author} {\bibfnamefont {V.}~\bibnamefont {Noel}}, \bibinfo {author} {\bibfnamefont {T.}~\bibnamefont {Gasenzer}},\ and\ \bibinfo {author} {\bibfnamefont {K.}~\bibnamefont {Boguslavski}},\ }\href {https://doi.org/10.1103/h4hn-r6kp} {\bibfield  {journal} {\bibinfo  {journal} {Phys. Rev. Res.}\ }\textbf {\bibinfo {volume} {7}},\ \bibinfo {pages} {033220} (\bibinfo {year} {2025})}\BibitemShut {NoStop}%
\bibitem [{\citenamefont {Rasch}\ \emph {et~al.}(2025)\citenamefont {Rasch}, \citenamefont {Chomaz},\ and\ \citenamefont {Gasenzer}}]{Rasch:2025kna}%
  \BibitemOpen
  \bibfield  {author} {\bibinfo {author} {\bibfnamefont {N.}~\bibnamefont {Rasch}}, \bibinfo {author} {\bibfnamefont {L.}~\bibnamefont {Chomaz}},\ and\ \bibinfo {author} {\bibfnamefont {T.}~\bibnamefont {Gasenzer}},\ }\href {https://doi.org/10.1103/x2rj-ptgy} {\bibfield  {journal} {\bibinfo  {journal} {Phys. Rev. A}\ }\textbf {\bibinfo {volume} {112}},\ \bibinfo {pages} {053310} (\bibinfo {year} {2025})}\BibitemShut {NoStop}%
\bibitem [{\citenamefont {Berges}(2016)}]{Berges2015a}%
  \BibitemOpen
  \bibfield  {author} {\bibinfo {author} {\bibfnamefont {J.}~\bibnamefont {Berges}},\ }in\ \href {https://doi.org/10.1093/acprof:oso/9780198768166.001.0001} {\emph {\bibinfo {booktitle} {Proc. Int. School on Strongly Interacting Quantum Systems Out of Equilibrium, Les Houches}}},\ \bibinfo {editor} {edited by\ \bibinfo {editor} {\bibfnamefont {T.}~\bibnamefont {{Giamarchi et al.}}}}\ (\bibinfo  {publisher} {OUP, Oxford},\ \bibinfo {year} {2016})\ pp.\ \bibinfo {pages} {69--206}\BibitemShut {NoStop}%
\bibitem [{\citenamefont {Berges}\ \emph {et~al.}(2021)\citenamefont {Berges}, \citenamefont {Heller}, \citenamefont {Mazeliauskas},\ and\ \citenamefont {Venugopalan}}]{Berges:2020fwq}%
  \BibitemOpen
  \bibfield  {author} {\bibinfo {author} {\bibfnamefont {J.}~\bibnamefont {Berges}}, \bibinfo {author} {\bibfnamefont {M.~P.}\ \bibnamefont {Heller}}, \bibinfo {author} {\bibfnamefont {A.}~\bibnamefont {Mazeliauskas}},\ and\ \bibinfo {author} {\bibfnamefont {R.}~\bibnamefont {Venugopalan}},\ }\href {https://doi.org/10.1103/RevModPhys.93.035003} {\bibfield  {journal} {\bibinfo  {journal} {Rev. Mod. Phys.}\ }\textbf {\bibinfo {volume} {93}},\ \bibinfo {pages} {035003} (\bibinfo {year} {2021})}\BibitemShut {NoStop}%
\bibitem [{\citenamefont {Mikheev}\ \emph {et~al.}(2023)\citenamefont {Mikheev}, \citenamefont {Siovitz},\ and\ \citenamefont {Gasenzer}}]{Mikheev2023a.EPJST232.3393}%
  \BibitemOpen
  \bibfield  {author} {\bibinfo {author} {\bibfnamefont {A.~N.}\ \bibnamefont {Mikheev}}, \bibinfo {author} {\bibfnamefont {I.}~\bibnamefont {Siovitz}},\ and\ \bibinfo {author} {\bibfnamefont {T.}~\bibnamefont {Gasenzer}},\ }\href {https://doi.org/10.1140/epjs/s11734-023-00974-7} {\bibfield  {journal} {\bibinfo  {journal} {Eur. Phys. J. Spec. Top.}\ }\textbf {\bibinfo {volume} {232}},\ \bibinfo {pages} {3393} (\bibinfo {year} {2023})}\BibitemShut {NoStop}%
\bibitem [{\citenamefont {Siovitz}\ \emph {et~al.}(2022)\citenamefont {Siovitz}, \citenamefont {Heinen}, \citenamefont {Rasch}, \citenamefont {Lannig}, \citenamefont {Deller}, \citenamefont {Strobel}, \citenamefont {Oberthaler},\ and\ \citenamefont {Gasenzer}}]{Siovitz2023b}%
  \BibitemOpen
  \bibfield  {author} {\bibinfo {author} {\bibfnamefont {I.}~\bibnamefont {Siovitz}}, \bibinfo {author} {\bibfnamefont {P.}~\bibnamefont {Heinen}}, \bibinfo {author} {\bibfnamefont {N.}~\bibnamefont {Rasch}}, \bibinfo {author} {\bibfnamefont {S.}~\bibnamefont {Lannig}}, \bibinfo {author} {\bibfnamefont {Y.}~\bibnamefont {Deller}}, \bibinfo {author} {\bibfnamefont {H.}~\bibnamefont {Strobel}}, \bibinfo {author} {\bibfnamefont {M.}~\bibnamefont {Oberthaler}},\ and\ \bibinfo {author} {\bibfnamefont {T.}~\bibnamefont {Gasenzer}},\ }in\ \href {https://doi.org/10.18419/opus-14626} {\emph {\bibinfo {booktitle} {Proceedings of the 8th bwHPC Symposium}}}\ (\bibinfo  {publisher} {OPUS},\ \bibinfo {year} {2022})\ pp.\ \bibinfo {pages} {51--56}\BibitemShut {NoStop}%
\bibitem [{\citenamefont {{Eigen}}\ \emph {et~al.}(2018)\citenamefont {{Eigen}}, \citenamefont {{Glidden}}, \citenamefont {{Lopes}}, \citenamefont {{Cornell}}, \citenamefont {{Smith}},\ and\ \citenamefont {{Hadzibabic}}}]{Eigen2018a}%
  \BibitemOpen
  \bibfield  {author} {\bibinfo {author} {\bibfnamefont {C.}~\bibnamefont {{Eigen}}}, \bibinfo {author} {\bibfnamefont {J.~A.~P.}\ \bibnamefont {{Glidden}}}, \bibinfo {author} {\bibfnamefont {R.}~\bibnamefont {{Lopes}}}, \bibinfo {author} {\bibfnamefont {E.~A.}\ \bibnamefont {{Cornell}}}, \bibinfo {author} {\bibfnamefont {R.~P.}\ \bibnamefont {{Smith}}},\ and\ \bibinfo {author} {\bibfnamefont {Z.}~\bibnamefont {{Hadzibabic}}},\ }\href {https://doi.org/10.1038/s41586-018-0674-1} {\bibfield  {journal} {\bibinfo  {journal} {Nature}\ }\textbf {\bibinfo {volume} {563}},\ \bibinfo {pages} {221} (\bibinfo {year} {2018})}\BibitemShut {NoStop}%
\bibitem [{\citenamefont {Pr{\"u}fer}\ \emph {et~al.}(2018)\citenamefont {Pr{\"u}fer}, \citenamefont {Kunkel}, \citenamefont {Strobel}, \citenamefont {Lannig}, \citenamefont {Linnemann}, \citenamefont {Schmied}, \citenamefont {Berges}, \citenamefont {Gasenzer},\ and\ \citenamefont {Oberthaler}}]{Prufer2018a}%
  \BibitemOpen
  \bibfield  {author} {\bibinfo {author} {\bibfnamefont {M.}~\bibnamefont {Pr{\"u}fer}}, \bibinfo {author} {\bibfnamefont {P.}~\bibnamefont {Kunkel}}, \bibinfo {author} {\bibfnamefont {H.}~\bibnamefont {Strobel}}, \bibinfo {author} {\bibfnamefont {S.}~\bibnamefont {Lannig}}, \bibinfo {author} {\bibfnamefont {D.}~\bibnamefont {Linnemann}}, \bibinfo {author} {\bibfnamefont {C.-M.}\ \bibnamefont {Schmied}}, \bibinfo {author} {\bibfnamefont {J.}~\bibnamefont {Berges}}, \bibinfo {author} {\bibfnamefont {T.}~\bibnamefont {Gasenzer}},\ and\ \bibinfo {author} {\bibfnamefont {M.~K.}\ \bibnamefont {Oberthaler}},\ }\href {https://doi.org/10.1038/s41586-018-0659-0} {\bibfield  {journal} {\bibinfo  {journal} {Nature}\ }\textbf {\bibinfo {volume} {563}},\ \bibinfo {pages} {217} (\bibinfo {year} {2018})}\BibitemShut {NoStop}%
\bibitem [{\citenamefont {Erne}\ \emph {et~al.}(2018)\citenamefont {Erne}, \citenamefont {B{\"u}cker}, \citenamefont {Gasenzer}, \citenamefont {Berges},\ and\ \citenamefont {Schmiedmayer}}]{Erne2018b}%
  \BibitemOpen
  \bibfield  {author} {\bibinfo {author} {\bibfnamefont {S.}~\bibnamefont {Erne}}, \bibinfo {author} {\bibfnamefont {R.}~\bibnamefont {B{\"u}cker}}, \bibinfo {author} {\bibfnamefont {T.}~\bibnamefont {Gasenzer}}, \bibinfo {author} {\bibfnamefont {J.}~\bibnamefont {Berges}},\ and\ \bibinfo {author} {\bibfnamefont {J.}~\bibnamefont {Schmiedmayer}},\ }\href {https://doi.org/10.1038/s41586-018-0667-0} {\bibfield  {journal} {\bibinfo  {journal} {Nature}\ }\textbf {\bibinfo {volume} {563}},\ \bibinfo {pages} {225} (\bibinfo {year} {2018})}\BibitemShut {NoStop}%
\bibitem [{\citenamefont {Navon}\ \emph {et~al.}(2019)\citenamefont {Navon}, \citenamefont {Eigen}, \citenamefont {Zhang}, \citenamefont {Lopes}, \citenamefont {Gaunt}, \citenamefont {Fujimoto}, \citenamefont {Tsubota}, \citenamefont {Smith},\ and\ \citenamefont {Hadzibabic}}]{Navon2018a.Science.366.382}%
  \BibitemOpen
  \bibfield  {author} {\bibinfo {author} {\bibfnamefont {N.}~\bibnamefont {Navon}}, \bibinfo {author} {\bibfnamefont {C.}~\bibnamefont {Eigen}}, \bibinfo {author} {\bibfnamefont {J.}~\bibnamefont {Zhang}}, \bibinfo {author} {\bibfnamefont {R.}~\bibnamefont {Lopes}}, \bibinfo {author} {\bibfnamefont {A.~L.}\ \bibnamefont {Gaunt}}, \bibinfo {author} {\bibfnamefont {K.}~\bibnamefont {Fujimoto}}, \bibinfo {author} {\bibfnamefont {M.}~\bibnamefont {Tsubota}}, \bibinfo {author} {\bibfnamefont {R.~P.}\ \bibnamefont {Smith}},\ and\ \bibinfo {author} {\bibfnamefont {Z.}~\bibnamefont {Hadzibabic}},\ }\href {https://doi.org/10.1126/science.aau6103} {\bibfield  {journal} {\bibinfo  {journal} {Science}\ }\textbf {\bibinfo {volume} {366}},\ \bibinfo {pages} {382} (\bibinfo {year} {2019})}\BibitemShut {NoStop}%
\bibitem [{\citenamefont {Glidden}\ \emph {et~al.}(2021)\citenamefont {Glidden}, \citenamefont {Eigen}, \citenamefont {Dogra}, \citenamefont {Hilker}, \citenamefont {Smith},\ and\ \citenamefont {Hadzibabic}}]{Glidden:2020qmu}%
  \BibitemOpen
  \bibfield  {author} {\bibinfo {author} {\bibfnamefont {J.~A.~P.}\ \bibnamefont {Glidden}}, \bibinfo {author} {\bibfnamefont {C.}~\bibnamefont {Eigen}}, \bibinfo {author} {\bibfnamefont {L.~H.}\ \bibnamefont {Dogra}}, \bibinfo {author} {\bibfnamefont {T.~A.}\ \bibnamefont {Hilker}}, \bibinfo {author} {\bibfnamefont {R.~P.}\ \bibnamefont {Smith}},\ and\ \bibinfo {author} {\bibfnamefont {Z.}~\bibnamefont {Hadzibabic}},\ }\href {https://doi.org/10.1038/s41567-020-01114-x} {\bibfield  {journal} {\bibinfo  {journal} {Nat. Phys.}\ }\textbf {\bibinfo {volume} {17}},\ \bibinfo {pages} {457} (\bibinfo {year} {2021})}\BibitemShut {NoStop}%
\bibitem [{\citenamefont {Garc\'{\i}a-Orozco}\ \emph {et~al.}(2022)\citenamefont {Garc\'{\i}a-Orozco}, \citenamefont {Madeira}, \citenamefont {Moreno-Armijos}, \citenamefont {Fritsch}, \citenamefont {Tavares}, \citenamefont {Castilho}, \citenamefont {Cidrim}, \citenamefont {Roati},\ and\ \citenamefont {Bagnato}}]{GarciaOrozco2021aNoArxiv}%
  \BibitemOpen
  \bibfield  {author} {\bibinfo {author} {\bibfnamefont {A.~D.}\ \bibnamefont {Garc\'{\i}a-Orozco}}, \bibinfo {author} {\bibfnamefont {L.}~\bibnamefont {Madeira}}, \bibinfo {author} {\bibfnamefont {M.~A.}\ \bibnamefont {Moreno-Armijos}}, \bibinfo {author} {\bibfnamefont {A.~R.}\ \bibnamefont {Fritsch}}, \bibinfo {author} {\bibfnamefont {P.~E.~S.}\ \bibnamefont {Tavares}}, \bibinfo {author} {\bibfnamefont {P.~C.~M.}\ \bibnamefont {Castilho}}, \bibinfo {author} {\bibfnamefont {A.}~\bibnamefont {Cidrim}}, \bibinfo {author} {\bibfnamefont {G.}~\bibnamefont {Roati}},\ and\ \bibinfo {author} {\bibfnamefont {V.~S.}\ \bibnamefont {Bagnato}},\ }\href {https://doi.org/10.1103/PhysRevA.106.023314} {\bibfield  {journal} {\bibinfo  {journal} {Phys. Rev. A}\ }\textbf {\bibinfo {volume} {106}},\ \bibinfo {pages} {023314} (\bibinfo {year} {2022})}\BibitemShut {NoStop}%
\bibitem [{\citenamefont {Lannig}\ \emph {et~al.}(2023)\citenamefont {Lannig}, \citenamefont {Pr{\"u}fer}, \citenamefont {Deller}, \citenamefont {Siovitz}, \citenamefont {Dreher}, \citenamefont {Gasenzer}, \citenamefont {Strobel},\ and\ \citenamefont {Oberthaler}}]{Lannig:2023fzf}%
  \BibitemOpen
  \bibfield  {author} {\bibinfo {author} {\bibfnamefont {S.}~\bibnamefont {Lannig}}, \bibinfo {author} {\bibfnamefont {M.}~\bibnamefont {Pr{\"u}fer}}, \bibinfo {author} {\bibfnamefont {Y.}~\bibnamefont {Deller}}, \bibinfo {author} {\bibfnamefont {I.}~\bibnamefont {Siovitz}}, \bibinfo {author} {\bibfnamefont {J.}~\bibnamefont {Dreher}}, \bibinfo {author} {\bibfnamefont {T.}~\bibnamefont {Gasenzer}}, \bibinfo {author} {\bibfnamefont {H.}~\bibnamefont {Strobel}},\ and\ \bibinfo {author} {\bibfnamefont {M.~K.}\ \bibnamefont {Oberthaler}}\ }\href {https://doi.org/10.48550/arXiv.2306.16497} {10.48550/arXiv.2306.16497} (\bibinfo {year} {2023}),\ \Eprint {https://arxiv.org/abs/2306.16497} {arXiv:2306.16497 [cond-mat.quant-gas]} \BibitemShut {NoStop}%
\bibitem [{\citenamefont {Martirosyan}\ \emph {et~al.}(2024)\citenamefont {Martirosyan}, \citenamefont {Ho}, \citenamefont {Etrych}, \citenamefont {Zhang}, \citenamefont {Cao}, \citenamefont {Hadzibabic},\ and\ \citenamefont {Eigen}}]{Martirosyan:2023mml}%
  \BibitemOpen
  \bibfield  {author} {\bibinfo {author} {\bibfnamefont {G.}~\bibnamefont {Martirosyan}}, \bibinfo {author} {\bibfnamefont {C.~J.}\ \bibnamefont {Ho}}, \bibinfo {author} {\bibfnamefont {J.}~\bibnamefont {Etrych}}, \bibinfo {author} {\bibfnamefont {Y.}~\bibnamefont {Zhang}}, \bibinfo {author} {\bibfnamefont {A.}~\bibnamefont {Cao}}, \bibinfo {author} {\bibfnamefont {Z.}~\bibnamefont {Hadzibabic}},\ and\ \bibinfo {author} {\bibfnamefont {C.}~\bibnamefont {Eigen}},\ }\href {https://doi.org/10.1103/PhysRevLett.132.113401} {\bibfield  {journal} {\bibinfo  {journal} {Phys. Rev. Lett.}\ }\textbf {\bibinfo {volume} {132}},\ \bibinfo {pages} {113401} (\bibinfo {year} {2024})},\ \Eprint {https://arxiv.org/abs/2304.06697} {arXiv:2304.06697 [cond-mat.quant-gas]} \BibitemShut {NoStop}%
\bibitem [{\citenamefont {Gazo}\ \emph {et~al.}(2025)\citenamefont {Gazo}, \citenamefont {Karailiev}, \citenamefont {Satoor}, \citenamefont {Eigen}, \citenamefont {Gałka},\ and\ \citenamefont {Hadzibabic}}]{Gazo2025a}%
  \BibitemOpen
  \bibfield  {author} {\bibinfo {author} {\bibfnamefont {M.}~\bibnamefont {Gazo}}, \bibinfo {author} {\bibfnamefont {A.}~\bibnamefont {Karailiev}}, \bibinfo {author} {\bibfnamefont {T.}~\bibnamefont {Satoor}}, \bibinfo {author} {\bibfnamefont {C.}~\bibnamefont {Eigen}}, \bibinfo {author} {\bibfnamefont {M.}~\bibnamefont {Gałka}},\ and\ \bibinfo {author} {\bibfnamefont {Z.}~\bibnamefont {Hadzibabic}},\ }\href {https://doi.org/10.1126/science.ado3487} {\bibfield  {journal} {\bibinfo  {journal} {Science}\ }\textbf {\bibinfo {volume} {389}},\ \bibinfo {pages} {802} (\bibinfo {year} {2025})}\BibitemShut {NoStop}%
\bibitem [{\citenamefont {Moreno-Armijos}\ \emph {et~al.}(2025)\citenamefont {Moreno-Armijos}, \citenamefont {Fritsch}, \citenamefont {Garc\'{\i}a-Orozco}, \citenamefont {Sab}, \citenamefont {Telles}, \citenamefont {Zhu}, \citenamefont {Madeira}, \citenamefont {Nazarenko}, \citenamefont {Yukalov},\ and\ \citenamefont {Bagnato}}]{MorenoArmijos2024a}%
  \BibitemOpen
  \bibfield  {author} {\bibinfo {author} {\bibfnamefont {M.~A.}\ \bibnamefont {Moreno-Armijos}}, \bibinfo {author} {\bibfnamefont {A.~R.}\ \bibnamefont {Fritsch}}, \bibinfo {author} {\bibfnamefont {A.~D.}\ \bibnamefont {Garc\'{\i}a-Orozco}}, \bibinfo {author} {\bibfnamefont {S.}~\bibnamefont {Sab}}, \bibinfo {author} {\bibfnamefont {G.}~\bibnamefont {Telles}}, \bibinfo {author} {\bibfnamefont {Y.}~\bibnamefont {Zhu}}, \bibinfo {author} {\bibfnamefont {L.}~\bibnamefont {Madeira}}, \bibinfo {author} {\bibfnamefont {S.}~\bibnamefont {Nazarenko}}, \bibinfo {author} {\bibfnamefont {V.~I.}\ \bibnamefont {Yukalov}},\ and\ \bibinfo {author} {\bibfnamefont {V.~S.}\ \bibnamefont {Bagnato}},\ }\href {https://doi.org/10.1103/PhysRevLett.134.023401} {\bibfield  {journal} {\bibinfo  {journal} {Phys. Rev. Lett.}\ }\textbf {\bibinfo {volume} {134}},\ \bibinfo {pages} {023401} (\bibinfo {year} {2025})}\BibitemShut {NoStop}%
\bibitem [{\citenamefont {Martirosyan}\ \emph {et~al.}(2025)\citenamefont {Martirosyan}, \citenamefont {Gazo}, \citenamefont {Etrych}, \citenamefont {Fischer}, \citenamefont {Morris}, \citenamefont {Ho}, \citenamefont {Eigen},\ and\ \citenamefont {Hadzibabic}}]{Martirosyan2024a}%
  \BibitemOpen
  \bibfield  {author} {\bibinfo {author} {\bibfnamefont {G.}~\bibnamefont {Martirosyan}}, \bibinfo {author} {\bibfnamefont {M.}~\bibnamefont {Gazo}}, \bibinfo {author} {\bibfnamefont {J.}~\bibnamefont {Etrych}}, \bibinfo {author} {\bibfnamefont {S.~M.}\ \bibnamefont {Fischer}}, \bibinfo {author} {\bibfnamefont {S.~J.}\ \bibnamefont {Morris}}, \bibinfo {author} {\bibfnamefont {C.~J.}\ \bibnamefont {Ho}}, \bibinfo {author} {\bibfnamefont {C.}~\bibnamefont {Eigen}},\ and\ \bibinfo {author} {\bibfnamefont {Z.}~\bibnamefont {Hadzibabic}},\ }\href {https://doi.org/10.1038/s41586-025-09735-z} {\bibfield  {journal} {\bibinfo  {journal} {Nature}\ }\textbf {\bibinfo {volume} {647}},\ \bibinfo {pages} {608} (\bibinfo {year} {2025})}\BibitemShut {NoStop}%
\bibitem [{\citenamefont {Zhu}\ \emph {et~al.}(2023)\citenamefont {Zhu}, \citenamefont {Xie},\ and\ \citenamefont {Xia}}]{Zhu2023a.PhysRevLett.130.214001}%
  \BibitemOpen
  \bibfield  {author} {\bibinfo {author} {\bibfnamefont {H.-Y.}\ \bibnamefont {Zhu}}, \bibinfo {author} {\bibfnamefont {J.-H.}\ \bibnamefont {Xie}},\ and\ \bibinfo {author} {\bibfnamefont {K.-Q.}\ \bibnamefont {Xia}},\ }\href {https://doi.org/10.1103/PhysRevLett.130.214001} {\bibfield  {journal} {\bibinfo  {journal} {Phys. Rev. Lett.}\ }\textbf {\bibinfo {volume} {130}},\ \bibinfo {pages} {214001} (\bibinfo {year} {2023})}\BibitemShut {NoStop}%
\bibitem [{\citenamefont {M\"uller}\ and\ \citenamefont {Krstulovic}(2024)}]{Mueller2024.PhysRevLett.132.094002}%
  \BibitemOpen
  \bibfield  {author} {\bibinfo {author} {\bibfnamefont {N.~P.}\ \bibnamefont {M\"uller}}\ and\ \bibinfo {author} {\bibfnamefont {G.}~\bibnamefont {Krstulovic}},\ }\href {https://doi.org/10.1103/PhysRevLett.132.094002} {\bibfield  {journal} {\bibinfo  {journal} {Phys. Rev. Lett.}\ }\textbf {\bibinfo {volume} {132}},\ \bibinfo {pages} {094002} (\bibinfo {year} {2024})}\BibitemShut {NoStop}%
\bibitem [{Note1()}]{Note1}%
  \BibitemOpen
  \bibinfo {note} {In classical turbulence, this exponent is usually denoted as $\xi $, with $\xi /2=\beta $.}\BibitemShut {Stop}%
\bibitem [{Note2()}]{Note2}%
  \BibitemOpen
  \bibinfo {note} {The term \protect \emph {anomalous} follows the RG convention, indicating that the scaling exponent deviates from its canonical value predicted by dimensional analysis.}\BibitemShut {Stop}%
\bibitem [{NoteTempQuenches()}]{NoteTempQuenches}%
  \BibitemOpen
  \bibinfo {note} {{In dissipative superfluids such as $^{4}$He, quenched in temperature, depending on boundary conditions and temperature, exponents $4\gtrsim z = \beta^{-1}\geq2$ have been reported \cite{Chu2001a,Forrester2013a.PhysRevLett.110.165303,Forrester2014a.JPhysConfSer568.012031,Forrester2020a,Williams2022a}.}}\BibitemShut {Stop}%
\bibitem [{\citenamefont {Benzi}\ \emph {et~al.}(1984)\citenamefont {Benzi}, \citenamefont {Paladin}, \citenamefont {Parisi},\ and\ \citenamefont {Vulpiani}}]{Benzi1984a}%
  \BibitemOpen
  \bibfield  {author} {\bibinfo {author} {\bibfnamefont {R.}~\bibnamefont {Benzi}}, \bibinfo {author} {\bibfnamefont {G.}~\bibnamefont {Paladin}}, \bibinfo {author} {\bibfnamefont {G.}~\bibnamefont {Parisi}},\ and\ \bibinfo {author} {\bibfnamefont {A.}~\bibnamefont {Vulpiani}},\ }\href {https://doi.org/10.1088/0305-4470/17/18/021} {\bibfield  {journal} {\bibinfo  {journal} {J. Phys. A}\ }\textbf {\bibinfo {volume} {17}},\ \bibinfo {pages} {3521} (\bibinfo {year} {1984})}\BibitemShut {NoStop}%
\bibitem [{Note3()}]{Note3}%
  \BibitemOpen
  \bibinfo {note} {{The derivation of intermittency from first principles remains an open problem in classical turbulence theory, cf., e.g., \cite{Zhou2021a.PhysRep935.1}.}}\BibitemShut {Stop}%
\bibitem [{\citenamefont {Boffetta}\ \emph {et~al.}(2000)\citenamefont {Boffetta}, \citenamefont {Celani},\ and\ \citenamefont {Vergassola}}]{Boffetta2000a}%
  \BibitemOpen
  \bibfield  {author} {\bibinfo {author} {\bibfnamefont {G.}~\bibnamefont {Boffetta}}, \bibinfo {author} {\bibfnamefont {A.}~\bibnamefont {Celani}},\ and\ \bibinfo {author} {\bibfnamefont {M.}~\bibnamefont {Vergassola}},\ }\href {https://doi.org/10.1103/PhysRevE.61.R29} {\bibfield  {journal} {\bibinfo  {journal} {Phys. Rev. E}\ }\textbf {\bibinfo {volume} {61}},\ \bibinfo {pages} {R29} (\bibinfo {year} {2000})}\BibitemShut {NoStop}%
\bibitem [{\citenamefont {Paret}\ and\ \citenamefont {Tabeling}(1998)}]{Paret1998a}%
  \BibitemOpen
  \bibfield  {author} {\bibinfo {author} {\bibfnamefont {J.}~\bibnamefont {Paret}}\ and\ \bibinfo {author} {\bibfnamefont {P.}~\bibnamefont {Tabeling}},\ }\href {https://doi.org/10.1063/1.869840} {\bibfield  {journal} {\bibinfo  {journal} {Phys. Fluids}\ }\textbf {\bibinfo {volume} {10}},\ \bibinfo {pages} {3126} (\bibinfo {year} {1998})}\BibitemShut {NoStop}%
\bibitem [{\citenamefont {Belin}\ \emph {et~al.}(1996)\citenamefont {Belin}, \citenamefont {Tabeling},\ and\ \citenamefont {Willaime}}]{Belin1996a.PhysicaD93.52}%
  \BibitemOpen
  \bibfield  {author} {\bibinfo {author} {\bibfnamefont {F.}~\bibnamefont {Belin}}, \bibinfo {author} {\bibfnamefont {P.}~\bibnamefont {Tabeling}},\ and\ \bibinfo {author} {\bibfnamefont {H.}~\bibnamefont {Willaime}},\ }\href {https://doi.org/10.1016/0167-2789(95)00279-0} {\bibfield  {journal} {\bibinfo  {journal} {Physica D}\ }\textbf {\bibinfo {volume} {93}},\ \bibinfo {pages} {52} (\bibinfo {year} {1996})}\BibitemShut {NoStop}%
\bibitem [{\citenamefont {Migdal}(1994)}]{Migdal1994a}%
  \BibitemOpen
  \bibfield  {author} {\bibinfo {author} {\bibfnamefont {A.~A.}\ \bibnamefont {Migdal}},\ }\href {https://doi.org/10.1142/S0217751X94000558} {\bibfield  {journal} {\bibinfo  {journal} {Int. J. Mod. Phys. A}\ }\textbf {\bibinfo {volume} {09}},\ \bibinfo {pages} {1197} (\bibinfo {year} {1994})}\BibitemShut {NoStop}%
\bibitem [{\citenamefont {Blakie}\ \emph {et~al.}(2008)\citenamefont {Blakie}, \citenamefont {Bradley}, \citenamefont {Davis}, \citenamefont {Ballagh},\ and\ \citenamefont {Gardiner}}]{Blakie2008a}%
  \BibitemOpen
  \bibfield  {author} {\bibinfo {author} {\bibfnamefont {P.~B.}\ \bibnamefont {Blakie}}, \bibinfo {author} {\bibfnamefont {A.~S.}\ \bibnamefont {Bradley}}, \bibinfo {author} {\bibfnamefont {M.~J.}\ \bibnamefont {Davis}}, \bibinfo {author} {\bibfnamefont {R.~J.}\ \bibnamefont {Ballagh}},\ and\ \bibinfo {author} {\bibfnamefont {C.~W.}\ \bibnamefont {Gardiner}},\ }\href {https://doi.org/10.1080/00018730802564254} {\bibfield  {journal} {\bibinfo  {journal} {Adv. Phys.}\ }\textbf {\bibinfo {volume} {57}},\ \bibinfo {pages} {363} (\bibinfo {year} {2008})}\BibitemShut {NoStop}%
%%CITATION = COND-MAT/0809.1487;%%
\bibitem [{SM()}]{SM}%
  \BibitemOpen
  \bibinfo {note} {{See Supplemental Material at \href{http://link.aps.org/supplemental/10.1103/p6wv-z621}{http://link.aps.org/\allowbreak supplemental/10.1103/p6wv-z621} for further details on the theoretical model, the characterization of universal dynamics close to the anomalous NTFP in a 2d Bose gas, and the statistics of velocity circulation, including the extraction of its universal scaling properties. It includes Refs.~\cite{Nore1997a.PhysRevLett.78.3896,Thudiyangal2024}.}}\BibitemShut {Stop}%
\bibitem [{Vid()}]{VideosNTFP2dTurb}%
  \BibitemOpen
  \href@noop {} {}\bibinfo {note} {{See video material at this \href{https://www.kip.uni-heidelberg.de/gasenzer/projects/decayingsfturbulence\#start}{https://www.kip.uni-heidelberg.de/gasenzer/projects/decayingsfturbulence\#start}.}}\BibitemShut {Stop}%
\bibitem [{\citenamefont {Heinen}\ \emph {et~al.}(2023)\citenamefont {Heinen}, \citenamefont {Mikheev},\ and\ \citenamefont {Gasenzer}}]{Heinen2023a}%
  \BibitemOpen
  \bibfield  {author} {\bibinfo {author} {\bibfnamefont {P.}~\bibnamefont {Heinen}}, \bibinfo {author} {\bibfnamefont {A.~N.}\ \bibnamefont {Mikheev}},\ and\ \bibinfo {author} {\bibfnamefont {T.}~\bibnamefont {Gasenzer}},\ }\href {https://doi.org/10.1103/PhysRevA.107.043303} {\bibfield  {journal} {\bibinfo  {journal} {Phys. Rev. A}\ }\textbf {\bibinfo {volume} {107}},\ \bibinfo {pages} {043303} (\bibinfo {year} {2023})}\BibitemShut {NoStop}%
\bibitem [{\citenamefont {Zhou}(2021)}]{Zhou2021a.PhysRep935.1}%
  \BibitemOpen
  \bibfield  {author} {\bibinfo {author} {\bibfnamefont {Y.}~\bibnamefont {Zhou}},\ }\href {https://doi.org/https://doi.org/10.1016/j.physrep.2021.07.001} {\bibfield  {journal} {\bibinfo  {journal} {Phys. Rep.}\ }\textbf {\bibinfo {volume} {935}},\ \bibinfo {pages} {1} (\bibinfo {year} {2021})}\BibitemShut {NoStop}%
\bibitem [{\citenamefont {Nore}\ \emph {et~al.}(1997)\citenamefont {Nore}, \citenamefont {Abid},\ and\ \citenamefont {Brachet}}]{Nore1997a.PhysRevLett.78.3896}%
  \BibitemOpen
  \bibfield  {author} {\bibinfo {author} {\bibfnamefont {C.}~\bibnamefont {Nore}}, \bibinfo {author} {\bibfnamefont {M.}~\bibnamefont {Abid}},\ and\ \bibinfo {author} {\bibfnamefont {M.~E.}\ \bibnamefont {Brachet}},\ }\href {https://doi.org/10.1103/PhysRevLett.78.3896} {\bibfield  {journal} {\bibinfo  {journal} {Phys. Rev. Lett.}\ }\textbf {\bibinfo {volume} {78}},\ \bibinfo {pages} {3896} (\bibinfo {year} {1997})}\BibitemShut {NoStop}%
\bibitem [{\citenamefont {Thudiyangal}\ and\ \citenamefont {del Campo}(2024)}]{Thudiyangal2024}%
  \BibitemOpen
  \bibfield  {author} {\bibinfo {author} {\bibfnamefont {M.}~\bibnamefont {Thudiyangal}}\ and\ \bibinfo {author} {\bibfnamefont {A.}~\bibnamefont {del Campo}},\ }\href {https://doi.org/10.1103/PhysRevResearch.6.033152} {\bibfield  {journal} {\bibinfo  {journal} {Phys. Rev. Res.}\ }\textbf {\bibinfo {volume} {6}},\ \bibinfo {pages} {033152} (\bibinfo {year} {2024})}\BibitemShut {NoStop}%
\end{thebibliography}

\providecommand*\hyphen{-}
%

%==============================================================================
\end{document}